\documentclass[preprint2]{aastex61}

\pdfoutput=1 %for arXiv submission
\usepackage{amsmath,amstext}
\usepackage[T1]{fontenc}
\usepackage{apjfonts} 
\usepackage[figure,figure*]{hypcap}
\usepackage{nameref}
\usepackage{hyperref}
\usepackage{subfigure}
\shorttitle{Whistler waves in the inner heliosphere}
\shortauthors{Jagarlamudi et al.}

\usepackage{soul,xcolor}
\begin{document}
\title{Whistler waves and electron properties in the inner heliosphere: \textit{Helios} Observations}

\author{Vamsee Krishna Jagarlamudi}
\affil{LPC2E/CNRS, 3 Avenue de la Recherche Scientifique, 45071 Orl\'eans Cedex 2, France}
\affiliation{LESIA, Observatoire de Paris, Universit\'e PSL, CNRS, Sorbonne Universit\'e, Universit\'e de Paris, 5 place Jules Janssen, 92195 Meudon, France}

\author{Olga Alexandrova}
\affiliation{LESIA, Observatoire de Paris, Universit\'e PSL, CNRS, Sorbonne Universit\'e, Universit\'e de Paris, 5 place Jules Janssen, 92195 Meudon, France}

\author{Laura Ber$\check{\textrm{c}}$i$\check{\textrm{c}}$}
\affiliation{LESIA, Observatoire de Paris, Universit\'e PSL, CNRS, Sorbonne Universit\'e, Universit\'e de Paris, 5 place Jules Janssen, 92195 Meudon, France}
\affiliation{Physics and Astronomy Department, University of Florence, Via Giovanni Sansone 1, I-50019 Sesto Fiorentino, Italy}

\author{Thierry Dudok de Wit}
\affiliation{LPC2E/CNRS, 3 Avenue de la Recherche Scientifique, 45071 Orl\'eans Cedex 2, France}
\author{Vladimir Krasnoselskikh}
\affiliation{LPC2E/CNRS, 3 Avenue de la Recherche Scientifique, 45071 Orl\'eans Cedex 2, France}

\author{Milan Maksimovic}
\affiliation{LESIA, Observatoire de Paris, Universit\'e PSL, CNRS, Sorbonne Universit\'e, Universit\'e de Paris, 5 place Jules Janssen, 92195 Meudon, France}

\author{$\check{\textrm{S}}$t$\check{\textrm{e}}$p\'an $\check{\textrm{S}}$tver\'ak}
\affiliation{Astronomical Institute, Czech Academy of Sciences, CZ-14100 Prague, Czech Republic}
\affiliation{Institute of Atmospheric Physics, Czech Academy of Sciences, CZ-14100 Prague, Czech Republic}

\received{April 9, 2020}

\accepted{May 15, 2020}

\submitjournal{ApJ}

\correspondingauthor{Vamsee Krishna Jagarlamudi}
\email{vamsee-krishna.jagarlamudi@cnrs-orleans.fr}

\begin{abstract}
We present the analysis of narrow-band whistler wave signatures observed in the inner heliosphere (0.3 to 1 AU). 
These signatures are bumps in the spectral density in the 10-200 Hz frequency range of the AC magnetic field as measured by the search coil magnetometer on-board the \textit{Helios}~1 spacecraft. We show that the majority of whistler signatures are observed in the slow solar wind ($<500$ kms$^{-1}$)  and their occurrence increases with the radial distance (R), from $\sim3\%$ at 0.3 AU to $\sim10\%$
at 0.9 AU. In the fast solar wind ($>600$ kms$^{-1}$), whistler activity is significantly lower, whistler signatures start to appear for $R>0.6$ AU and their number increases from $\sim0.03\%$ at 0.65 AU to $\sim1\%$ at 0.9 AU.
We have studied the variation of the electron core and halo anisotropy ($T_{e\perp}/T_{e\parallel}$), as well as the electron normalized heat flux as a function of R and of the solar wind speed. We find that in the slow wind electron core and halo anisotropy is higher than in fast one, and also these anisotropies increase radially in both types of winds, which is in line with the occurrence of whistler signatures. We hypothesize the existence of a feedback mechanism to explain the observed radial variations
of the occurrence of whistlers in relation with the halo anisotropy.

\end{abstract}

\section{Introduction}
Whistler waves are the most probable electromagnetic modes observed between the lower hybrid frequency ($f_{LH}=\sqrt{f_{ce}f_{ci}}$) and electron cyclotron frequency $(f_{ce})$ \citep{Gary1993}, where $f_{ci}={q_pB}/{2\pi m_p}$, $f_{ce}={q_eB}/{2\pi m_e}$, $q_p$ is charge of the proton, $q_e$ is charge of the electron, $m_p$ is mass of the electron, $m_e$ is mass of the electron and $B$ is magnitude of the magnetic field. These waves are thought to have a significant contribution in the regulation of the fundamental processes in the solar wind, especially for the solar wind electrons. Whistler waves are therefore widely studied phenomena and are the key to a better understanding of the global solar wind thermodynamics and energy transport.

Whistlers could play a important role in the evolution of the solar wind electron velocity distributions \citep{Vocks2005,Vocks2012,Kajdic2016,Roberg2018,Boldyrev2019,Tang2020} through the pitch angle scattering of the beam-like solar wind electron component, called the strahl \citep{Hammond1996,Graham2017}, which is in turn is expected to affect the  second high energy electron population, the halo. Whistler waves are expected to regulate the heat flux \citep{Garry1994,Gary1999,Roberg2018}. On the other hand, kinetic simulations \citep{Vocks2003,Vocks2012} have shown that whistler waves could also provide a mechanism for the continuous formation of suprathermal electrons in the corona. In the Earth's magnetosphere, whistler waves have a significant role in the acceleration and precipitation of particles in the radiation belts, e.g. \citet{artemyev16}.

%%%%%%%%%%%%%%%%%%%%%%%%%%%%%%%%%%%%%%%%%%%%
%%%%%%%%%%%%%%%%%%%%%%%%%%%%%%%%%%%%%%%%%%
%%%%%%%%%%%%%%%%%%%%%%%%%%%%%%%%%%%%%%%%%%

In this study, we focus on whistler waves in the inner heliosphere (for radial distances between 0.3 and 1 AU).  
One of the earliest studies of magnetic fluctuations at kinetic scales in the inner-heliosphere was done by \citet{Beinroth1981}. The authors used the power spectral density of the magnetic field measured by the search coil magnetometer on-board \textit{Helios}~1. They interpreted the observed broadband spectrum as due to whistler waves. Later it became clear that these broadband spectra represent background turbulence  \citep{Alexandrova2012} with wave vectors mostly perpendicular to the mean field $\mathbf{B}$, $k_{\perp}\gg k_{\|}$ and negligible frequencies in the plasma frame \citep{Lacombe2017}. Instead, the whistler waves observed up to date in the free solar wind are narrow band, quasi-parallel, right handed polarized; in power spectral densities they appear as spectral bumps 
e.g., \citet{Lacombe2014,Kajdic2016,Roberts2017}. Note that the observed frequency range in the satellite frame is with in the expected frequency range in the plasma frame, namely, between the $f_{LH}$ and $f_{ce}$ \citep{Gary1993,Stenzel1999}.

Among the early studies which confirmed the existence of whistler waves in the solar wind was that by \citet{Zhang1998}. They used magnetic and electric field high resolution  waveform data on board  the \textit{Geotail} spacecraft upstream of the bow shock. In the free solar wind, the authors observed short-lived (less than 1~s) monochromatic wave-packets, at frequencies between $0.1f_{ce}$ and $0.4f_{ce}$ with RH polarisation and $\mathbf{k}$ mostly parallel to $\mathbf{B}$, in anti-sunward direction. 

An analysis of long-lived (5-10 minutes), narrow-band, quasi-parallel, RH  whistlers in the free solar wind was done by \citet{Lacombe2014} using the magnetic spectral matrix routine measurements of the \textit{Cluster}/STAFF instrument, between 2001 and 2005. About 20 such events were found in the slow wind, within the $[f_{LH},0.5f_{ce}]$ frequency range.

\citet{Stansby2016} studied the RH wavepackets observed in the whistler range using the \textit{Artemis} electric and magnetic field wave forms measured at 128 samples/second in the solar wind. The authors determined empirical dispersion relation and found a good agreement with the linear theory predictions for parallel propagating whistlers in a finite $\beta_e$ plasma. They showed that the observed whistlers are
aligned with the background magnetic field and majority of them travel anti-sunward.

\citet{Breneman2010J} studied high-resolution electric field data from \textit{Stereo/WAVES/TDS}, which capture the most intense signals. 
The authors found quasi-monochromatic, oblique, RH polarised whistlers around $\sim 0.1 f_{ce}$. These oblique whistler waves are mostly associated with the stream interaction regions (SIR's) and interplanetary (IP) shocks and have unusual high amplitudes in comparison with what was known before \citep{Moullard2001}. The duration of these waves is found being few seconds to minutes.

What are the sources of whistler waves ? The primary source of energy in the pristine solar wind are instabilities driven by the electron distribution functions.
In the frame of the solar wind electrons, we expect to see two types of instabilities: the whistler temperature anisotropy instability, which depends on the ratio between perpendicular and parallel electron temperatures $T_{e\perp} / T_{e\parallel}>1$, and the whistler heat flux instability is expected to develop when the heat flux is mainly carried by the anti-sunward motion of the halo/strahl electrons relative to the sunward moving core  \citep{Gary1977,Breneman2010J}.

\citet{Zhang1998} suggested that the parallel short-lived whistlers are generated by the halo electrons. In the data set of \citet{Lacombe2014}, when long-lived whistlers were observed the temperature anisotropy of the electrons was $T_{e\perp}/T_{e\|}<1$, while the heat flux showed a relative increase at the time of appearance of the waves, thus indicating that the whistler heat flux instability is probably responsible for the observed whistler activity. Note that the values of the temperatures were those of the global electron distribution function, without separation between the core, the halo and the strahl.

\citet{Wilson2013} investigated whistler waves downstream of super-critical interplanetary shocks and suggested that these waves might be driven by a heat flux instability and cause perpendicular heating of the halo electrons. 
\citet{Tong2019} studied the simultaneous field and particle measurements when the field aligned whistler waves are observed and they showed that the temperature anisotropy of halo electrons significantly affects the heat flux instability onset.

In this study, we focus on  signatures of  
whistler waves within the solar wind (excluding magnetic clouds and IP shocks) using search coil magnetometer (SCM) data from \textit{Helios}~1. 
Spanning the distances between 0.3 and 1 AU, the \textit{Helios} 1 mission allows us to investigate the radial dependence of the properties of these waves. The drawback of the used data set is that it only includes SCM power spectral density (PSD) of two components, and that the waveform
data were lost. Therefore, there is no information on the polarisation of the waves, which makes it impossible to identify whistler waves in an unambiguous way. 
We assume that any local concentration of spectral power (i.e. a spectral bump) observed in the frequency range $f_{lh}<f<f_{ce}$, are signatures of whistlers. Therefore, the presence of a spectral bump in the spectrum is our sole criterion for whistler wave detection.

The article is structured as follows. In Section~\ref{sec:data} we describe the \textit{Helios} 1 data used for our analysis. In Section~\ref{sec:Identification and conformation} we present the method of whistler wave identification. In Section~\ref{sec:Whistler wave properties} we present the properties of the observed whistler waves in the inner heliosphere. In Section~\ref{sec:Discussions} we discuss the possible mechanism which could explain the observed whistler properties and this is followed by the conclusions in Section~\ref{sec:conclusion}.

\section{Data }

\label{sec:data}

We use magnetic field spectral density data from the search coil magnetometer \citep{Dehmel1975} on-board \textit{Helios}~1 taken from the \href{http://helios-data.ssl.berkeley.edu/}{Helios Data Archive}. 
The data we use is the time series of the spectral density of the $B_y$ component of the sensor (located in the ecliptic plane) that were collected during the period of 12-12-1974 to 20-09-1975 with an 8-second cadence. For a comprehensive discussion on \textit{Helios} search coil magnetometer (SCM) and its data, we refer the reader to the \href{http://helios-data.ssl.berkeley.edu/}{Helios Data Archive}.

The spectral density of the $B_y$ component is measured in 8 logarithmically-spaced frequency bands. Their central frequencies are respectively : 6.8 Hz, 14.7 Hz, 31.6 Hz, 68 Hz, 147 Hz, 316 Hz, 681 Hz and 1470 Hz. From these we obtain the power spectral density (PSD) by squaring the spectral density. Although the $B_z$ component (perpendicular to the ecliptic plane) is also measured, it is more affected by the spacecraft electromagnetic noise \citep{Neubauer1977} and so we discard it. The SCM has two data products: mean and peak magnetic field spectral density in the considered time interval. Depending on the operational mode of spacecraft telemetry system, the mean and peak amplitudes are calculated for intervals of 1.125, 2.25, 4.5, 18, 36, 72, 144, 288, 576 or 1152 seconds. If the average interval is less than 4.5 s then the mean values are compressed to 8 second averages (\href{http://helios-data.ssl.berkeley.edu/}{Helios data archive}). Peak amplitudes frequently saturate near the Sun; for that reason we consider mean amplitudes only.

We consider the PSD's that exceed the SCM noise level by a factor of 2 at least for the first 4 bands. The SCM noise is taken from \citep{Neubauer1977}. We obtain 324366 spectra for our analysis.

For the physical interpretation of our data we use the proton and electron moments. Proton moments such as density, velocity and temperature of 40.5 s time resolution are taken from the \href{http://helios-data.ssl.berkeley.edu/}{Helios data archive}. The electron moments (01-01-1975 to 10-05-1979) are taken from the work of \citet{Stverak2009} and \citet{Stverak2015}, here authors have considered only the measurements when the magnetic field vector is almost in an ecliptic plane. Unfortunately these electron observations of 40.5 s time resolution are less regularly available.

\begin{figure}
\centering
\includegraphics[width=0.9\linewidth]{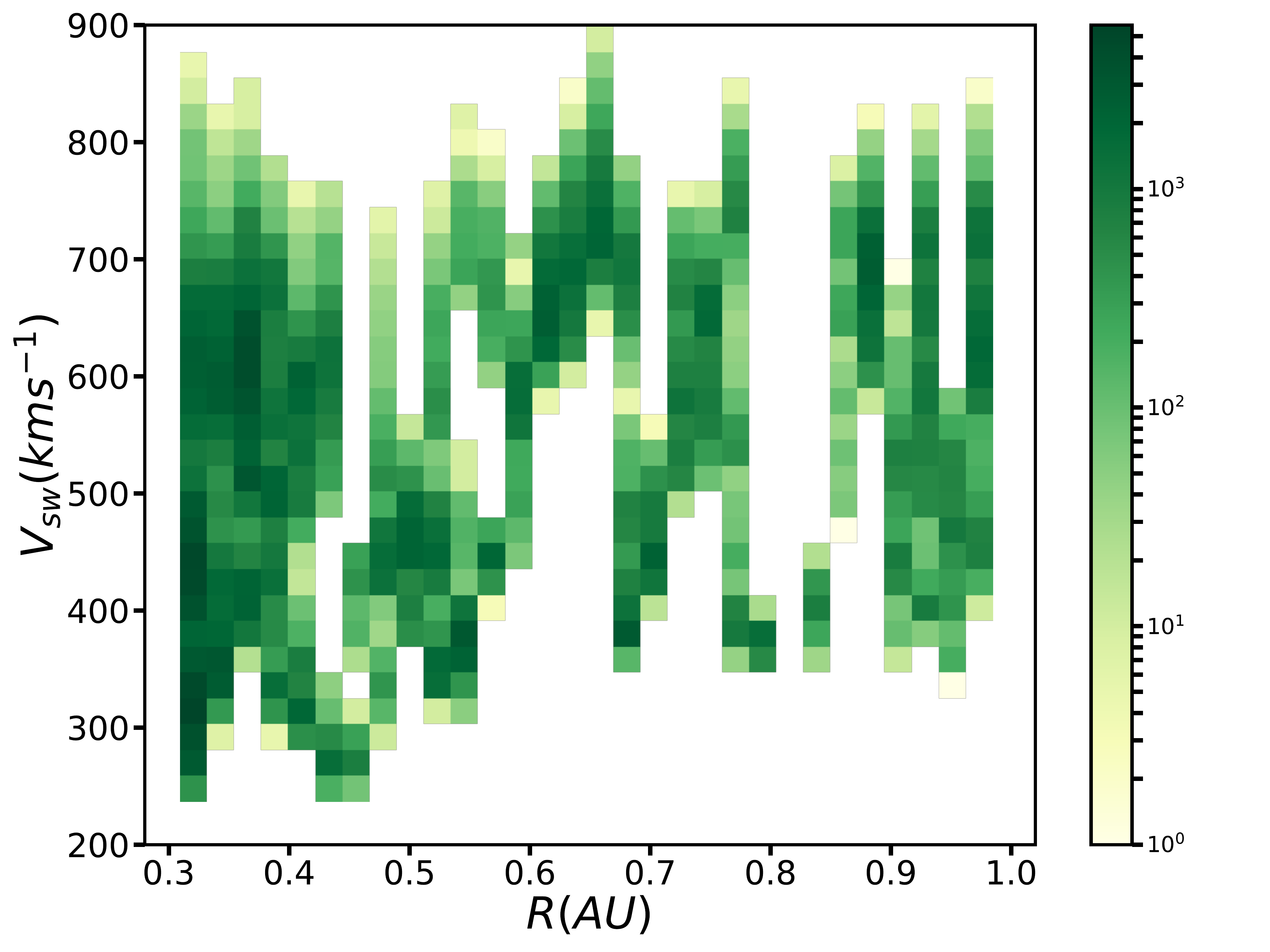}
%\decoRule
\caption{2d histogram of solar wind bulk speed ($V_{sw}$) corresponding to the analyzed spectra for the whole Helios-1 mission, from December 12, 1974 to September 20 1975 as a function of distance ($R$) from the Sun.}
\label{fig:VelocitySCM position}
\end{figure}

We divided the SCM magnetic spectra observations based on the solar wind bulk velocity $(V_{sw})$, fast solar wind ($V_{sw}>600$ kms$^{-1}$) and slow solar wind ($V_{sw}<500$ kms$^{-1}$). In Figure \ref{fig:VelocitySCM position} we show the 2D histogram of $V_{sw}$ for the analyzed data set as a function of the distance from the Sun $R$. We observe that the slow and fast wind SCM data are available at all distances (0.3 to 1 AU) for our analysis. There are nearly equal amounts of observations made in the slow and in the fast solar wind: 43 \% of the spectra are in the slow wind and 37 \% in the fast wind. The remaining 20 \% correspond to intermediate velocities.

\begin{figure}
\centering
\includegraphics[width=1.0\linewidth]{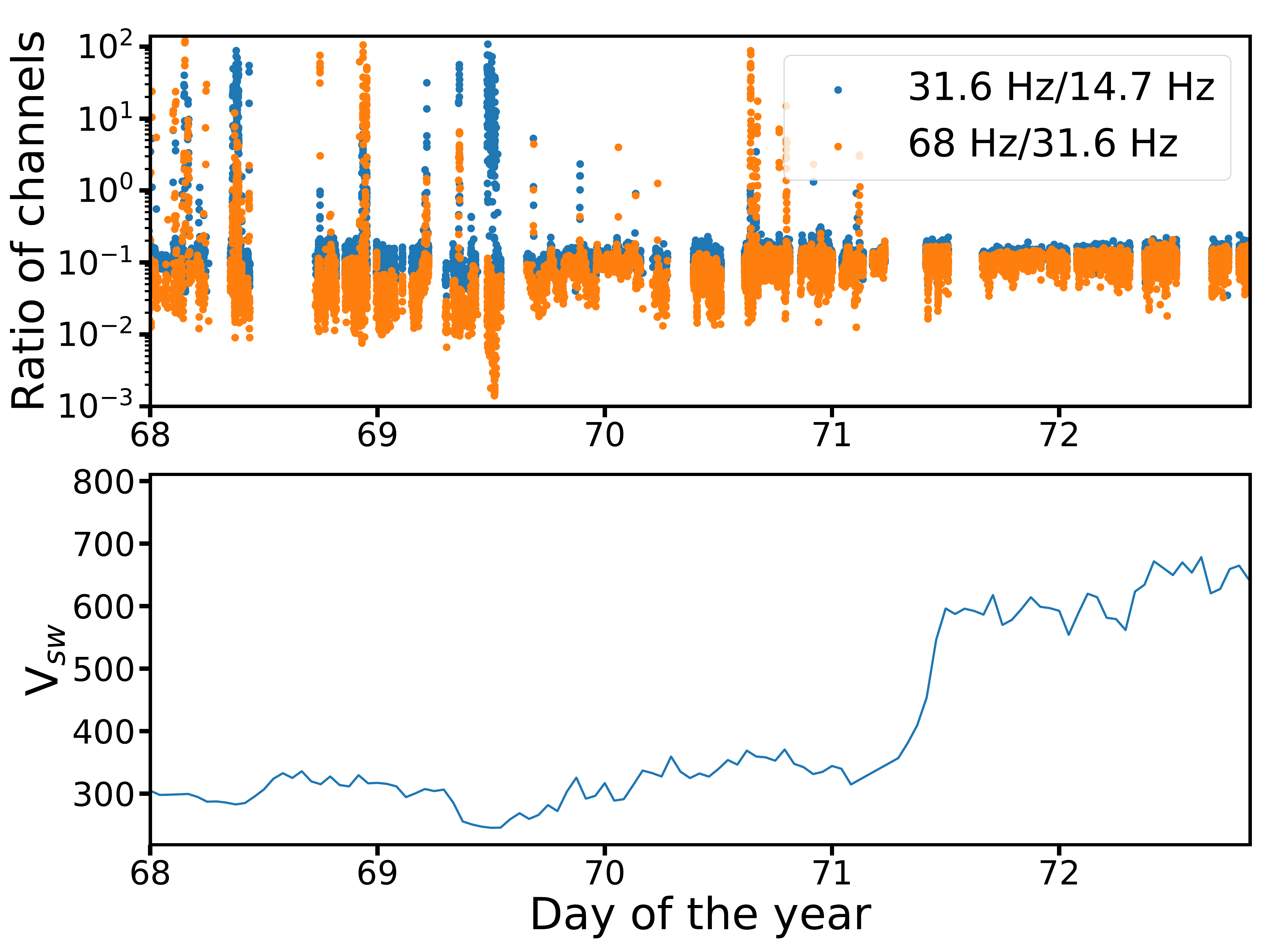}
%\decoRule
\caption{The ratio of PSDs between two pairs of frequency channels (upper panel) and variation of the solar wind velocity (lower panel) around 0.3 AU in 1975 from \textit{Helios} 1.}
\label{fig:channelratios}
\end{figure}

The general behavior of the slow and fast solar wind spectra is shown in Figure \ref{fig:channelratios} using the ratio of the amplitude of the PSD between different frequency channels as a function of time. In the fast wind, the ratio of amplitudes between two consecutive channels are almost constant and therefore nearly all the spectral bands show similar behavior. This coherent evolution of the spectra means that the spectral shape is conserved and only the amplitude of the spectrum is varying. A completely different picture emerges from the slow wind, in which the channel ratios show frequent intermittent bursts, which are signatures of large changes of the spectral shape.

\newpage

\section{Whistler wave identification}
\label{sec:Identification and conformation}

\begin{figure}%
\centering
\subfigure(a){%
\label{fig:first}%
\includegraphics[width=0.8\linewidth]{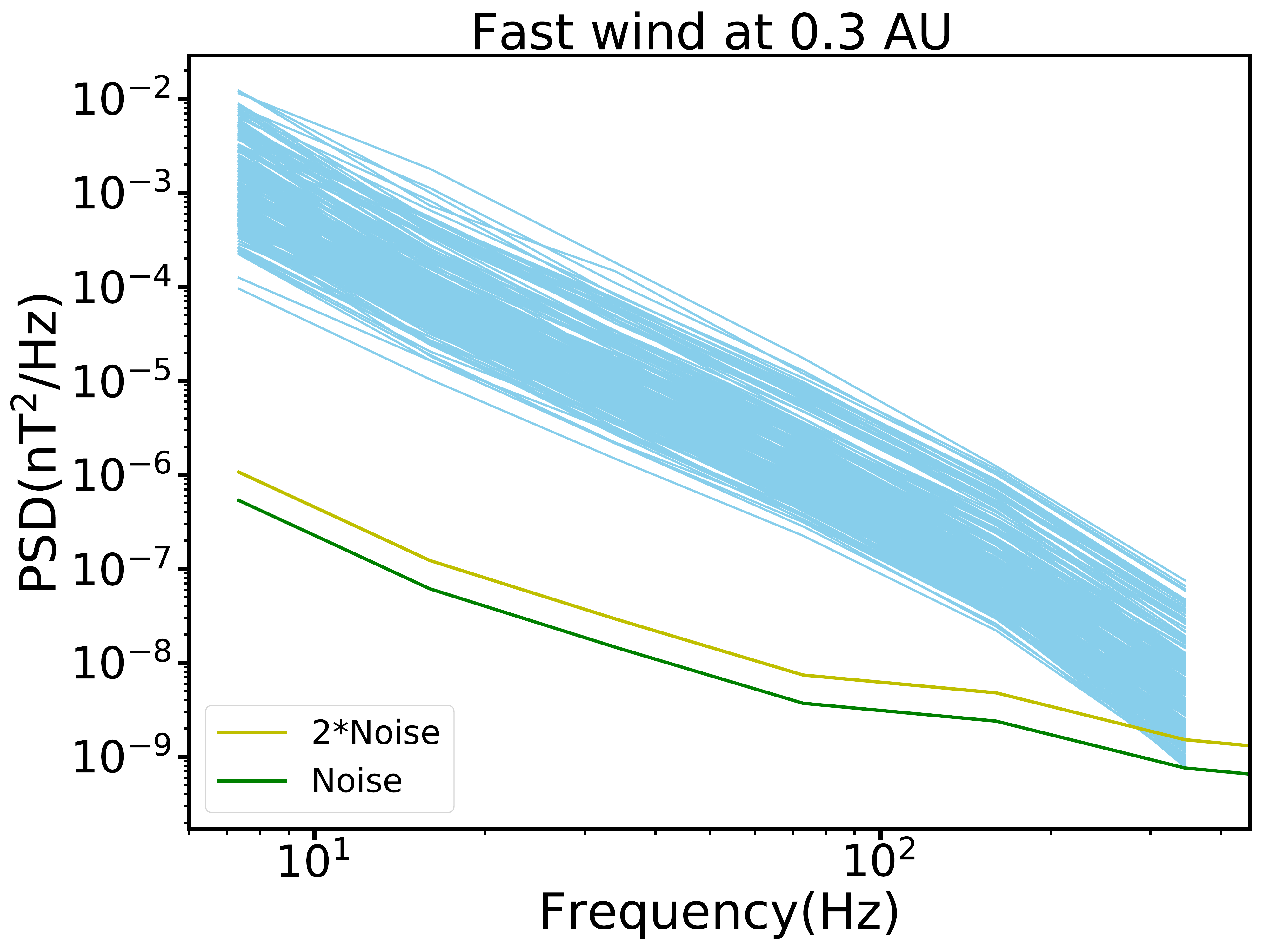}}%
\qquad
\subfigure(b){%
\label{fig:second}%
\includegraphics[width=0.8\linewidth]{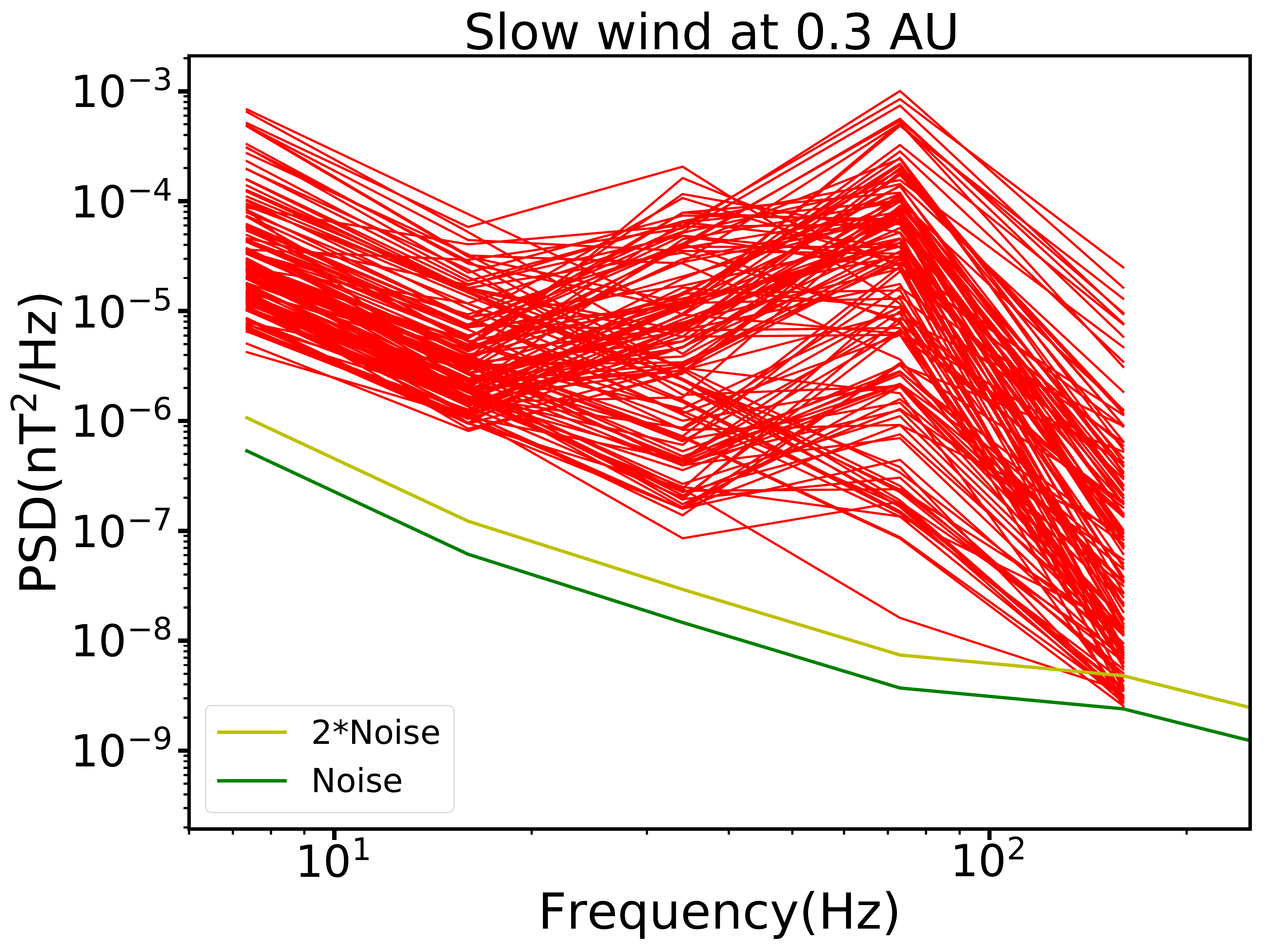}}%
\caption{Example of typical Helios1/SCM spectra at 0.3 AU. Panel (a) shows the spectra (217) in the fast wind interval (10:00-10:45, 9-03-1975); panel (b) shows the spectra (136) in the slow wind (19:45-20:30, 14-03-1975). }
\label{fig:Example spectra}
\end{figure}

Figure \ref{fig:Example spectra} illustrates typical spectra in the fast (panel (a)) and slow wind (panel (b)) at 0.3 AU. Here one observes:  
\begin{enumerate}
    \item Smooth spectra, where amplitude is decreasing monotonically with frequency (Figure \ref{fig:Example spectra}~(a)). This is a permanent feature of background magnetic field turbulence as was observed at 1~AU \citep{Alexandrova2012}.
     \item Spectra with local maxima (Figure \ref{fig:Example spectra}~(b)): background magnetic field turbulence spectra influenced by narrow-band
     %either long-lived or large amplitude quasi-monochromatic 
     waves as in \citep{Lacombe2014}.
    \item Some spectra which are in between, without a distinctive local maxima. Those are the ones that might have been affected by either short lived or very low amplitude waves.
\end{enumerate}

We consider those spectra in which one single local maximum (bump) clearly stands out with respect to the PSD of the background turbulence. Mathematically speaking, $\frac{dPSD}{df}$ is negative for class 1 spectra as we go towards higher frequencies, this kind of \textit{Helios} spectra are analysed in \citet{Alexandrova2020}. However, when the whistler waves influence the spectra, as we go towards higher frequencies at a certain frequency $\frac{dPSD}{df}$ will be positive and then automatically again go to negative as for class 2 spectra. In this way, we automatically identify the presence of spectral bump. Using the list of interplanetary shocks \citep{lucas2011,Kruparova2013} and the list of magnetic clouds \citep{Bothmer1998} we have eliminated the interplanetary shocks and magnetic cloud intervals from our analysis. In our total analyzed samples of $N_{tot}=$324366 we have observed $N_w=$7287 spectra that have a distinctive spectral bump. This number is small ($\sim 2.2 \%$) compared to the total number of spectra analyzed ($N_{tot}$). 
However, this number is comparable to the percentage of whistler waves observed at 1~AU by \citet{Tong2019stasticalstudy}, which is $\sim 1.7 \%$. Note that we miss sporadic or low amplitude whistler waves with our selection criteria.

The visibility of a clear bump in a PSD may depend on many factors such as lifetime, amplitude and frequency of the wave packet, on the bandwidth of the instrument and also on the level of turbulence at that moment.
We created synthetic data to approach this problem.
For this purpose we use an example from \citep{Lacombe2014}: we take the background turbulence level within the kinetic range and whistler wave packets at 30~Hz, with the amplitude about $\delta B/B\simeq 0.05$. 
We observe that a distinctive bump appears for waves that dominate the background fluctuation during the time of integration of the spectrum. 
However, even a lifetime of $1/16$ in the considered interval can be enough to observe a bump if the wave dominates the background.

%%%%%%%%%%%%%%%%%%%%%%
%%%%%%%%%%%%%%%%%%%%%%%%

%%%%%%%%%%%%%%%5
%%%%%%%%%%%%%%%%%%%%

The next step is the identification of the central frequency of the spectral bumps and to analyze whether the observed frequencies are in the whistler range or not.

\begin{figure}%
\centering
\subfigure(a){%
\label{fig:first}%
\includegraphics[width=0.8\linewidth]{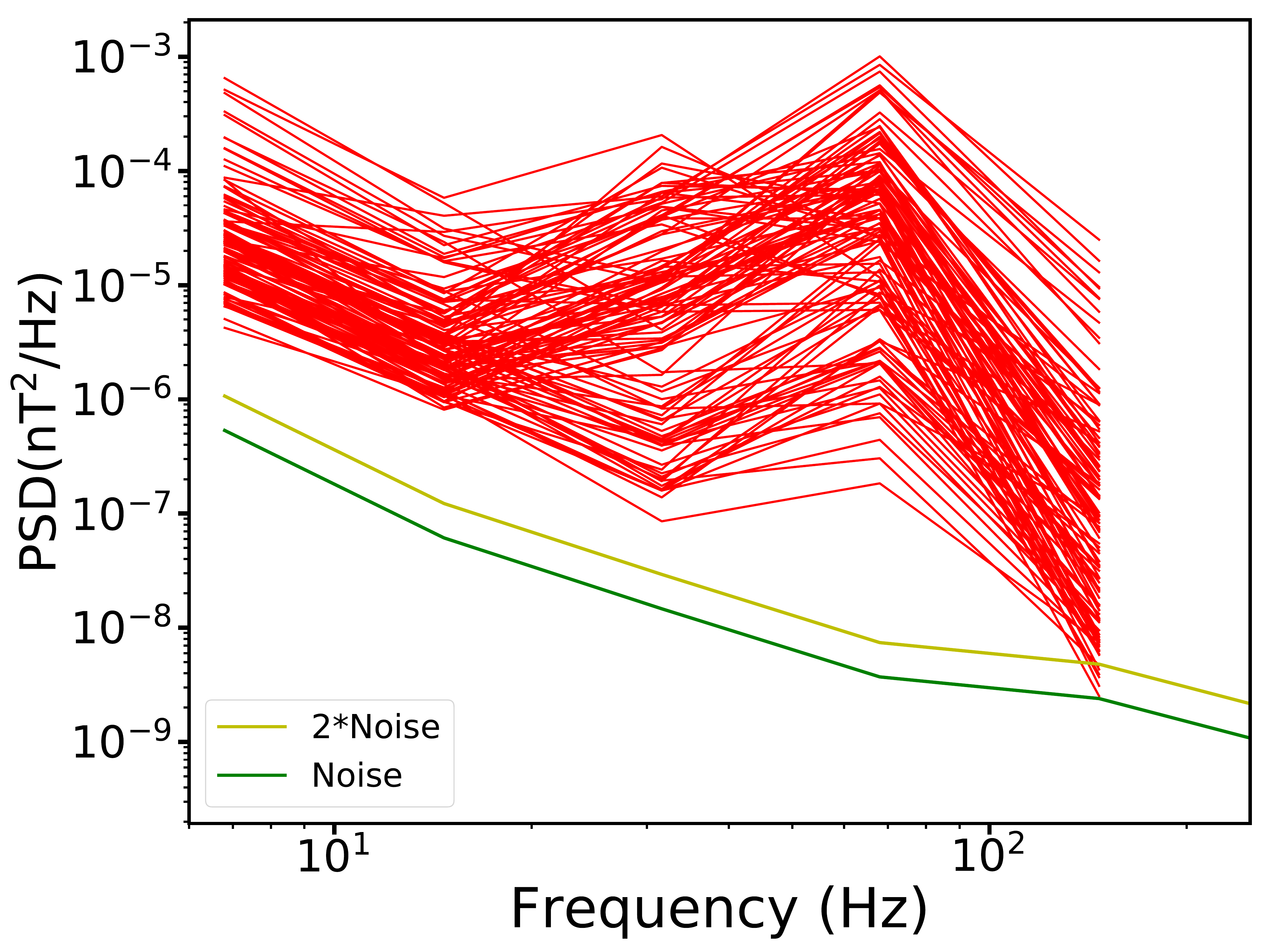}}%
\qquad
\subfigure(b){%
\label{fig:second}%
\includegraphics[width=0.8\linewidth]{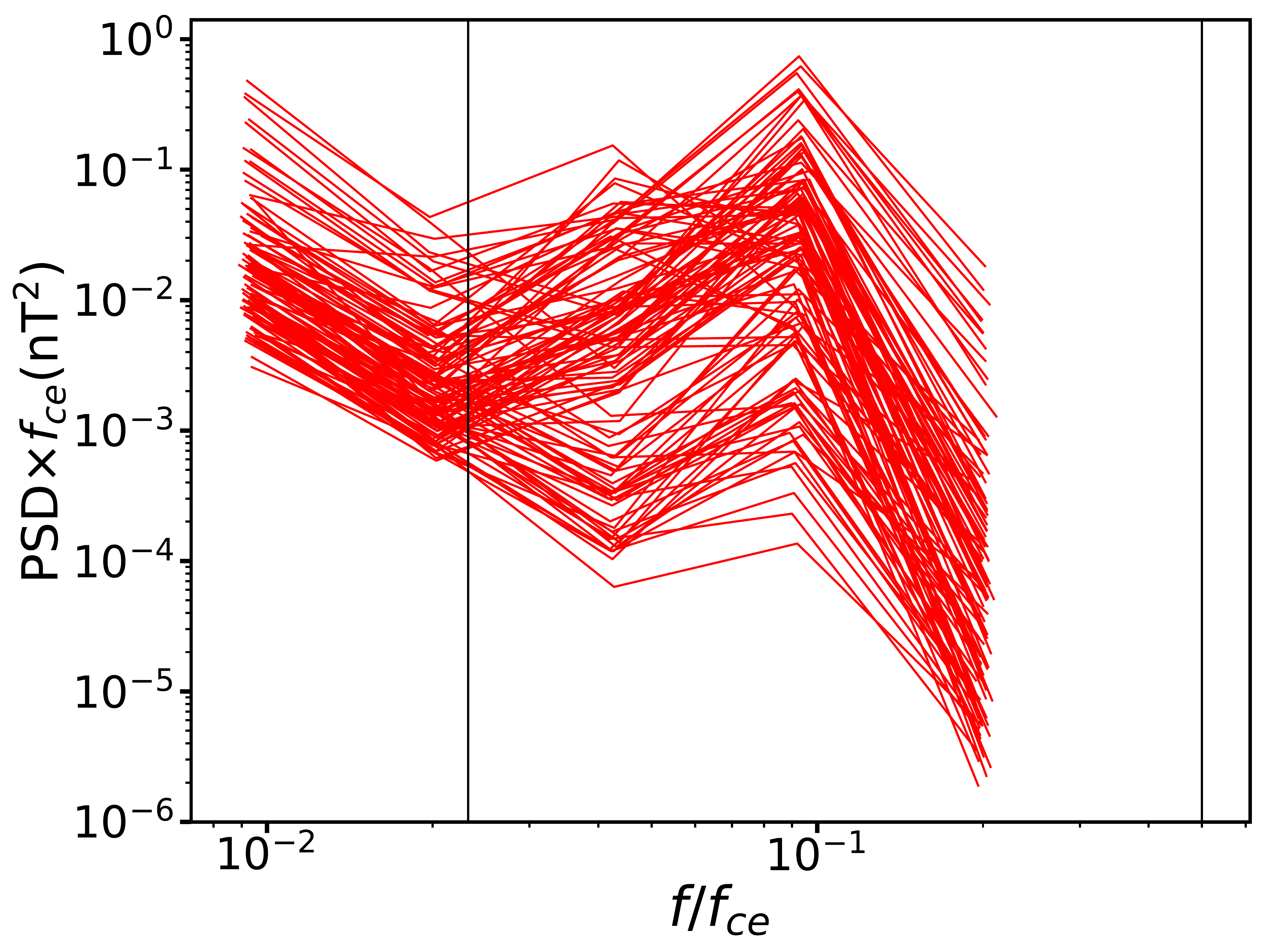}}%
\caption{Illustration of 115 spectra with clear bumps out of 136 spectra of Figure \ref{fig:Example spectra} (b) as a function of f in panel (a) and $f/f_{ce}$ in panel (b). The region between the black vertical lines correspond to whistler range.}
\label{fig:Example normalized spectra}
\end{figure}

Figure \ref{fig:Example normalized spectra} shows spectra with distinctive local maxima
extracted from the example spectra shown in Figure \ref{fig:Example spectra} (b). In Figure \ref{fig:Example normalized spectra} (a) we show the spectra as a function of the measured $f$  and in Figure \ref{fig:Example normalized spectra} (b) we show the same spectra with respect to ${f}/{f_{ce}}$. From Figure \ref{fig:Example normalized spectra} (b) we can infer that for the observed example spectra, the spectral bumps are in the whistler range, i.e. between $\sim f_{LH}$ and 0.5 $f_{ce}$, see two vertical lines. The mean frequency is around $f/f_{ce}=0.1$.
Now, let us verify the central frequency of the bumps for all the spectra.

\begin{figure}
\centering
\includegraphics[width=0.9\linewidth]{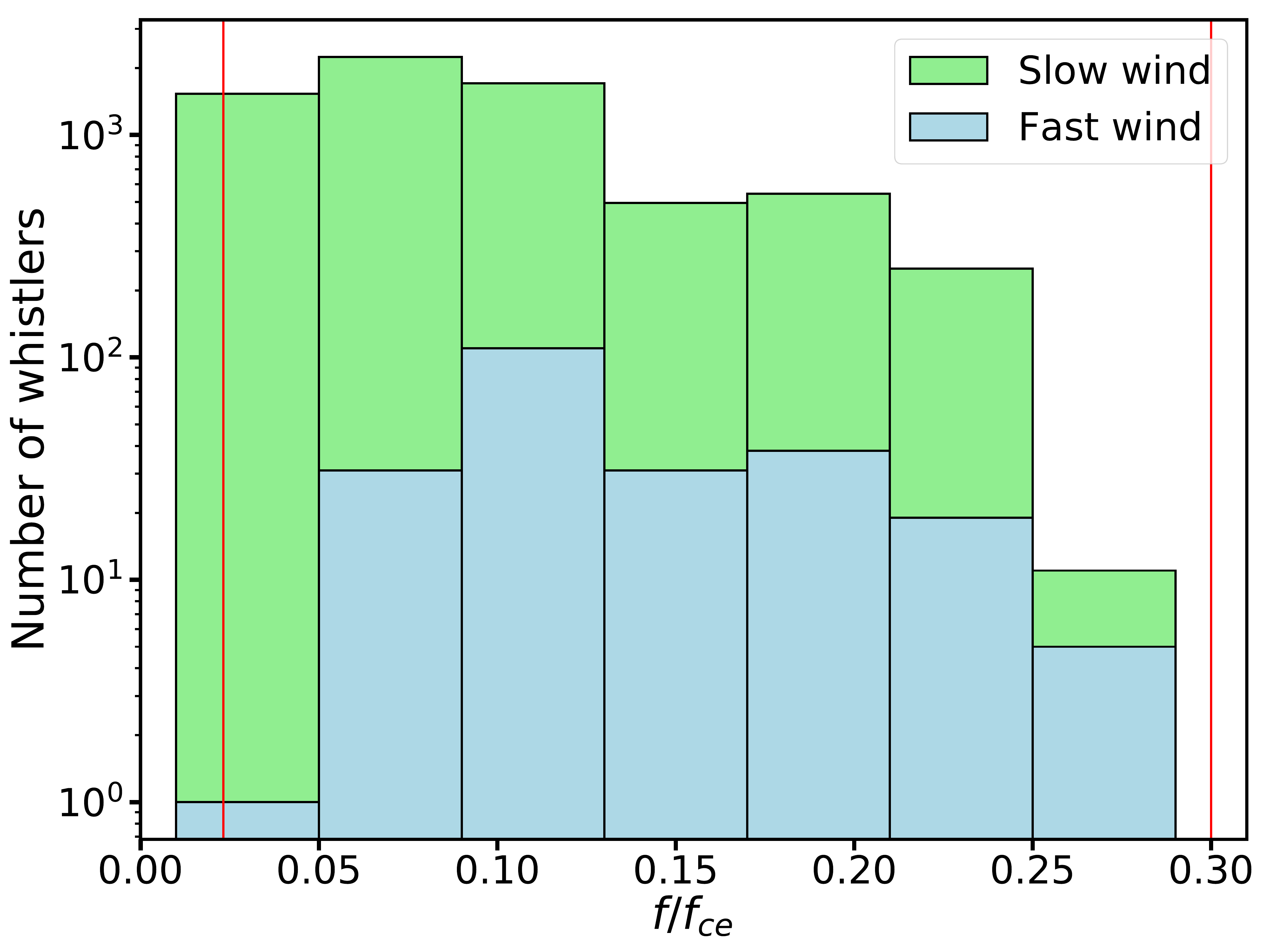}
%\decoRule
\caption{Histograms showing the distribution of normalized frequencies of the spectral bumps. Green histogram correspond to the slow wind, while the blue histogram correspond to the fast wind. The red vertical lines correspond to $f_{LH}/f_{ce}$ and 0.3 respectively.}
\label{fig:normalized frequency}
\end{figure}

In Figure \ref{fig:normalized frequency} we show the distribution of normalized central frequencies for the $N_w$ bumps observed in the Helios1/SCM spectra. Almost all of the bumps are between $\sim f_{LH}$ and  $0.3f_{ce}$. There were 12 spectra which were observed very close but below the whistler range, however, we have included them in the analysis, as Doppler shift may be the reason for that. 

As was discussed in the introduction, \citet{Lacombe2014} have shown that in the free solar wind any spectral bump in $[f_{LH},0.5 f_{ce}]$ range corresponds to right handed (RH), circularly polarized whistlers.
All the whistlers were observed to be propagating along the mean magnetic field. As far as we have no access to polarization properties, we use the results of \citet{Lacombe2014} as a basis for our assumption that any bump we observe at frequencies $f_{LH}<f<0.5f_{ce}$  corresponds to whistler waves.

\section{Whistler wave properties}
\label{sec:Whistler wave properties}

\begin{figure*}
\centering
\includegraphics[width=0.9\linewidth]{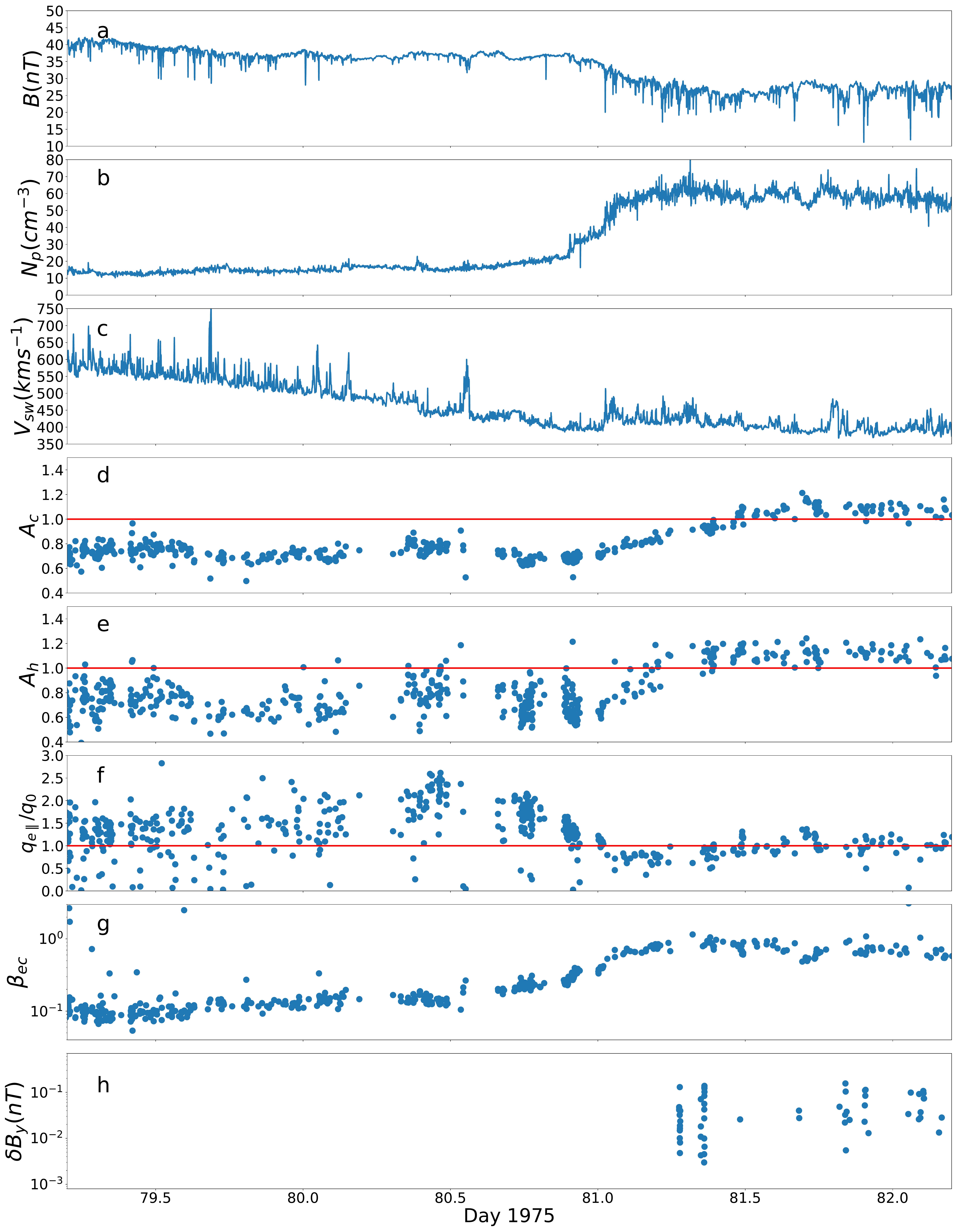}
%\decoRule
\caption{Illustration of typical conditions when the whistler waves are observed using a three day interval around 0.4 AU on March 21-24, 1975. We show in panel (a) the magnitude of magnetic field, in panel (b) we show the proton density, in panel (c) the proton bulk velocity, in panel (d) the electron core anisotropy ($A_{c}$) and the red line corresponds to $A_{c}=1$, in panel (e) the halo anisotropy ($A_{h}$) and the red line corresponds to $A_{h}=1$, in panel (f) the normalized parallel electron heat flux ($q_{e\parallel}/q_0$) and the red line corresponds to $q_{e\parallel}/q_0=1$, in panel (g) the electron core beta and in panel (h) the amplitudes of the intermittently occurring whistlers waves.}
\label{fig:plasma parameters}
\end{figure*}

In Figure \ref{fig:plasma parameters} we show an example of the typical plasma parameters around 0.4 AU along with the amplitudes of the observed whistler wave signatures. This figure gives us a glimpse of the plasma conditions when the whistler waves are observed. We infer that whistler waves occur intermittently and have different amplitudes (the definition of the amplitude of the waves is given in section \ref{sec:amplitude}). Whistler waves appear in the slow solar wind, where the typical conditions when compared to the fast solar wind are: lower magnetic field magnitude, higher proton density, higher electron core anisotropy ($A_{c}=\frac{T_{e\perp c}}{T_{e\parallel c}} $), higher electron halo anisotropy ($A_{h}=\frac{T_{e\perp h}}{T_{e\parallel h}} $), lower normalized heat flux ($q_{e \parallel}/q_0$), where $q_0=1.5n_eT_e(T_e/m_e)^{3/2}$ and higher $\beta_{ec}$ $(n_{ec}k_BT_{ec}/\frac{B^2}{2\mu_0})$. In the coming subsections, we shall discuss in detail the effects of these properties.    

%\newpage

\subsection{Solar wind bulk speed and whistler waves}

\begin{figure}
\centering
\includegraphics[width=0.9\linewidth]{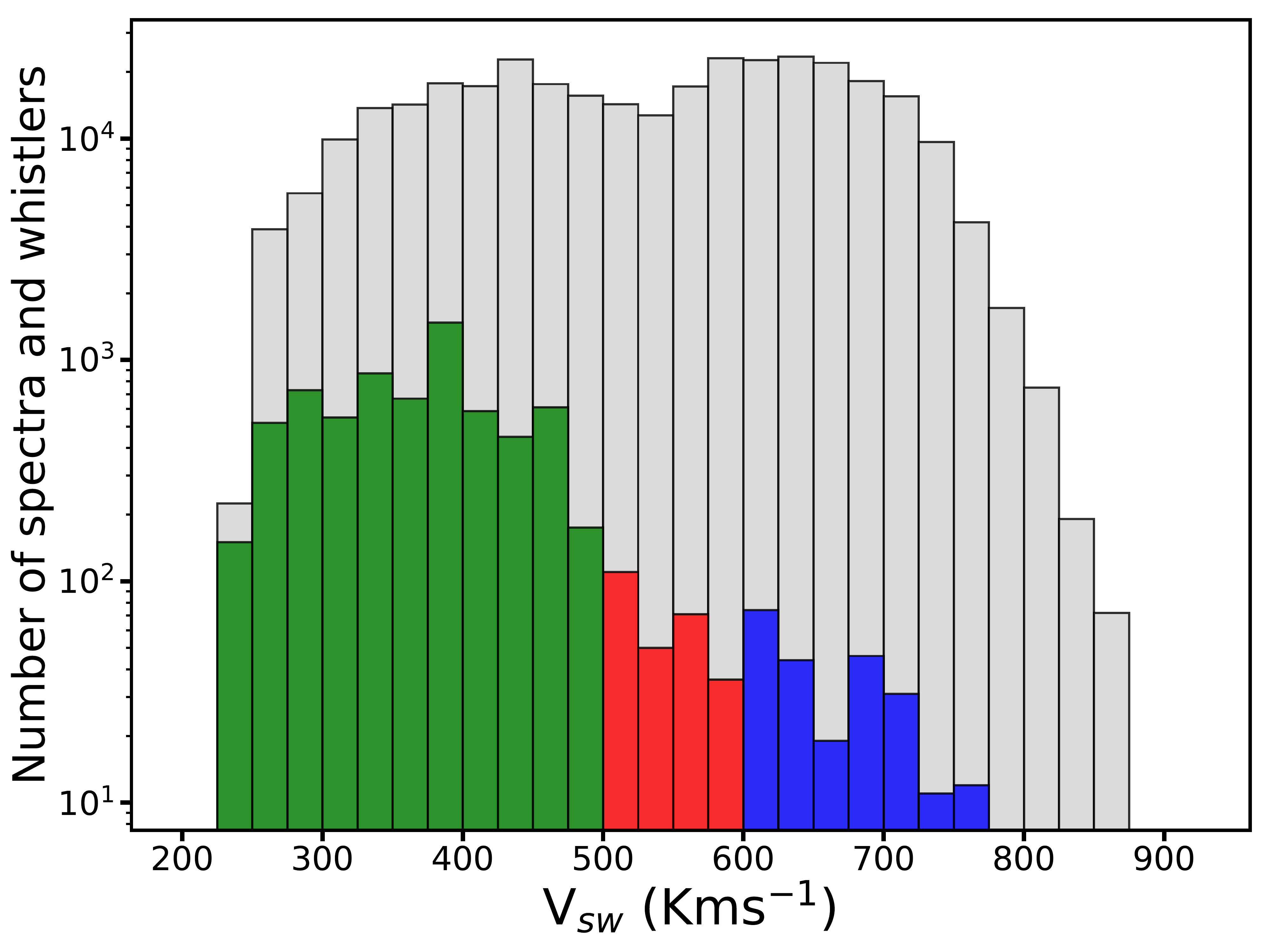}
%\decoRule
\caption{Distribution of the solar wind velocities for the spectra used in our analysis. The histogram of all analysed spectra is shown in grey, while green, red, and blue histograms show the fraction with identified whistler waves for slow, intermediate, and fast solar wind, respectively. }
\label{fig:VelocitySCM whistlers}
\end{figure}

Histograms in Figure~\ref{fig:VelocitySCM whistlers} show the distribution of solar wind velocities for all $N_{tot}$ spectra used in our analysis and for the spectra showing the presence of whistler wave signatures. All the analyzed spectra are shown in grey, slow wind whistlers are shown in green, intermediate in red and fast wind whistlers are shown in blue, respectively.

Even though we have observed nearly equal numbers of spectra in the slow and fast wind, the majority of the spectra with signatures of whistlers $\sim$ 93$\%$ (6783) were observed in the slow solar wind ( $<500$ kms$^{-1}$). This is similar to the study by \citet{Lacombe2014} at 1 AU, where the authors suggested that a slow wind speed is one of the necessary conditions for the observation of long-lived whistler waves. Interestingly, this condition at 1 AU also seems to hold in the inner heliosphere. However, in our study, we also observe whistler waves in the fast solar wind but they are rare, and represent about 3$\%$ (237) of  the number of the spectra with wave signatures. The remaining 4\% are observed in the intermediate wind.

\begin{figure}
\centering
\includegraphics[width=0.9\linewidth]{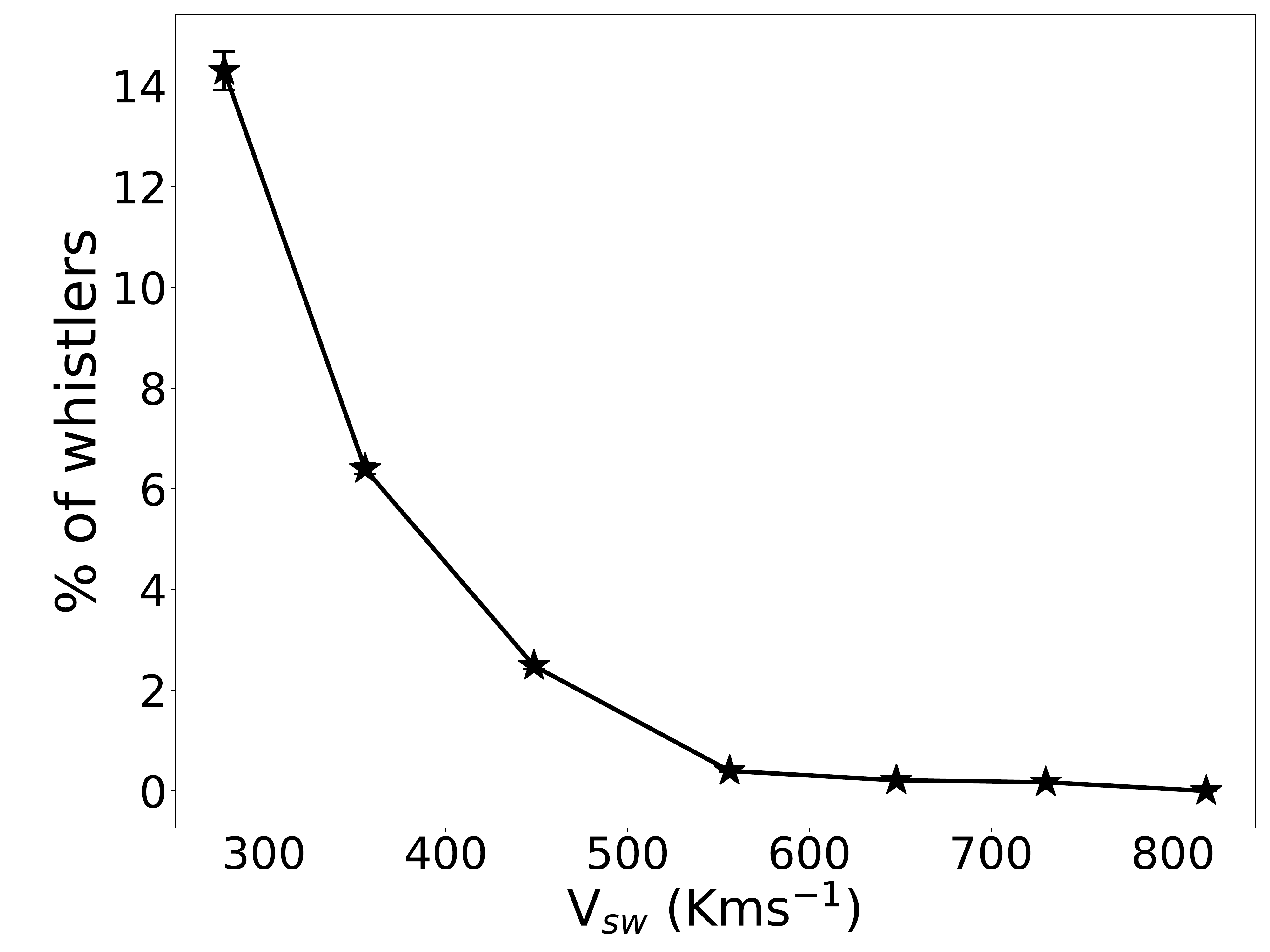}
%\decoRule
\caption{The occurrence of whistler waves as a function of the solar wind bulk speed ($V_{sw}$). For each velocity we show the fraction of PSDs that have the signature of whistler waves. Error bars correspond to standard error, i.e. $10^2\sqrt{N_{w,V_{sw}}}/N_{tot,V_{sw}}$.}
\label{fig:percentage whistlers velocitySCM}
\end{figure}

Figure \ref{fig:percentage whistlers velocitySCM} shows the percentage of whistler waves as a function of the wind velocity, i.e. the number of whistlers to the number of spectra available in the corresponding velocity bin. We observe that there is a constant decrease of whistler waves occurrence as the wind velocity is increasing : the higher the velocity, the lower the probability of observing whistler waves.

\subsection{Radial distribution of the observed whistler waves }

\begin{figure}
\centering
\includegraphics[width=0.9\linewidth]{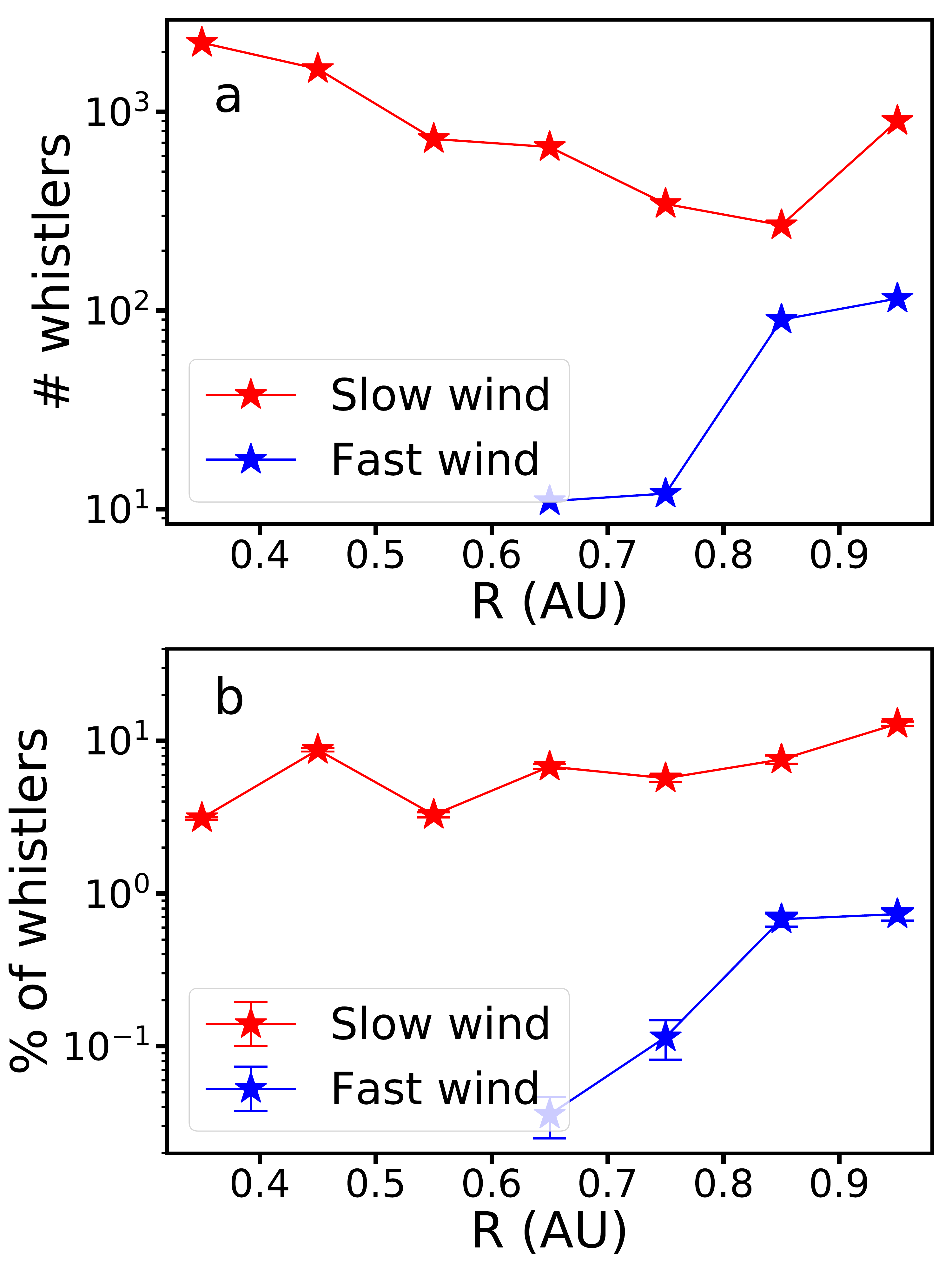}
%\decoRule
\caption{Radial variation of number of whistler waves and their occurrence in the slow (red) and fast (blue) wind as a function of distance from the Sun. Panel (a) shows the number of whistler waves observed in the slow and fast wind at different distances. Panel (b) shows the percentage of the slow wind whistlers with respect to the total number of the spectra in the slow wind for the corresponding distance bin and similarly for the fast wind. Error bars correspond to standard error, i.e. $10^2\sqrt{N_{w}}/N_{tot}$. }
\label{fig:whistler percentage}
\end{figure}

Figure \ref{fig:whistler percentage} (a) shows the number of spectra with whistler wave signatures observed at different radial distances from the Sun. We observe that in the slow wind whistler waves are present at all the distances, from 0.3 to 1 AU. In contrast, in the fast wind whistlers only start to appear for $R>0.6$~AU, in spite of large number of spectra available closer to the Sun (see Figure \ref{fig:VelocitySCM position}). 

Figure \ref{fig:whistler percentage} (b) shows the occurrence of whistlers in the slow and fast wind as a function of radial distance between 0.3 to 1 AU. Precisely, we show the percentage of spectra with signatures of whistler waves to the spectra analysed in the corresponding radial distance bin for the slow (red) and fast (blue) wind.

Surprisingly, we observe the percentage of whistler waves increases as we move away from the Sun in the slow and fast wind. However, the occurrence of whistlers in the fast wind is very low when compared to the slow wind. This is the first time that an increase in the percentage of whistlers with the radial distance has been shown.

\subsection{Amplitude of the observed whistler waves}
\label{sec:amplitude}
%%%%%%%%%%%%%%%%%%%%%%%%%%%%%%%

\begin{figure}
\centering
\includegraphics[width=0.9\linewidth]{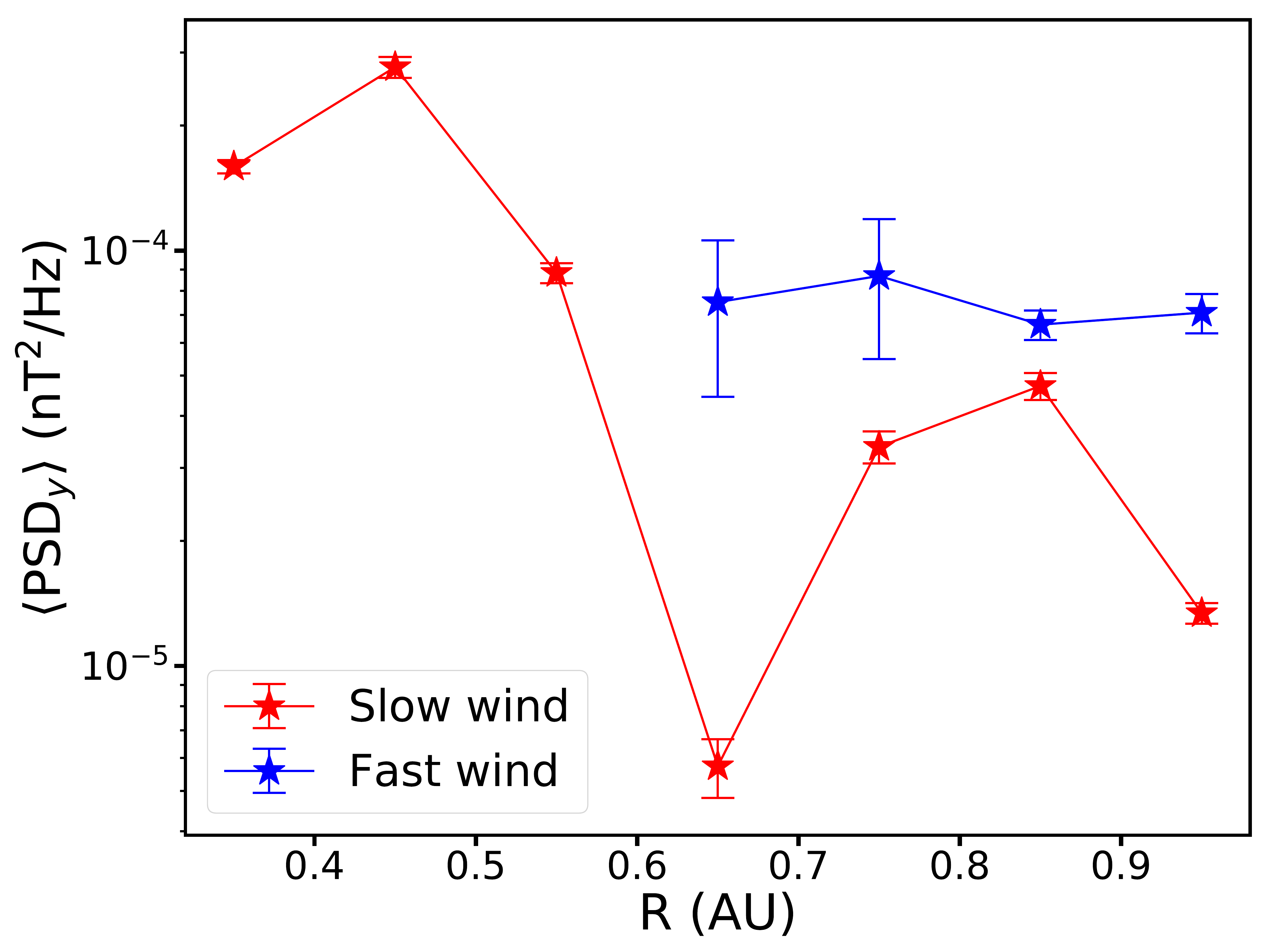}
%\decoRule
\caption{ Mean power spectral density of the peak value of the spectral bumps in the slow and fast wind at different distances from the Sun. Error bars show the standard error ($\frac{\sigma}{\sqrt{n}}$).}
\label{fig:psd_whistlers}
\end{figure}

\begin{figure}
\centering
\includegraphics[width=0.9\linewidth]{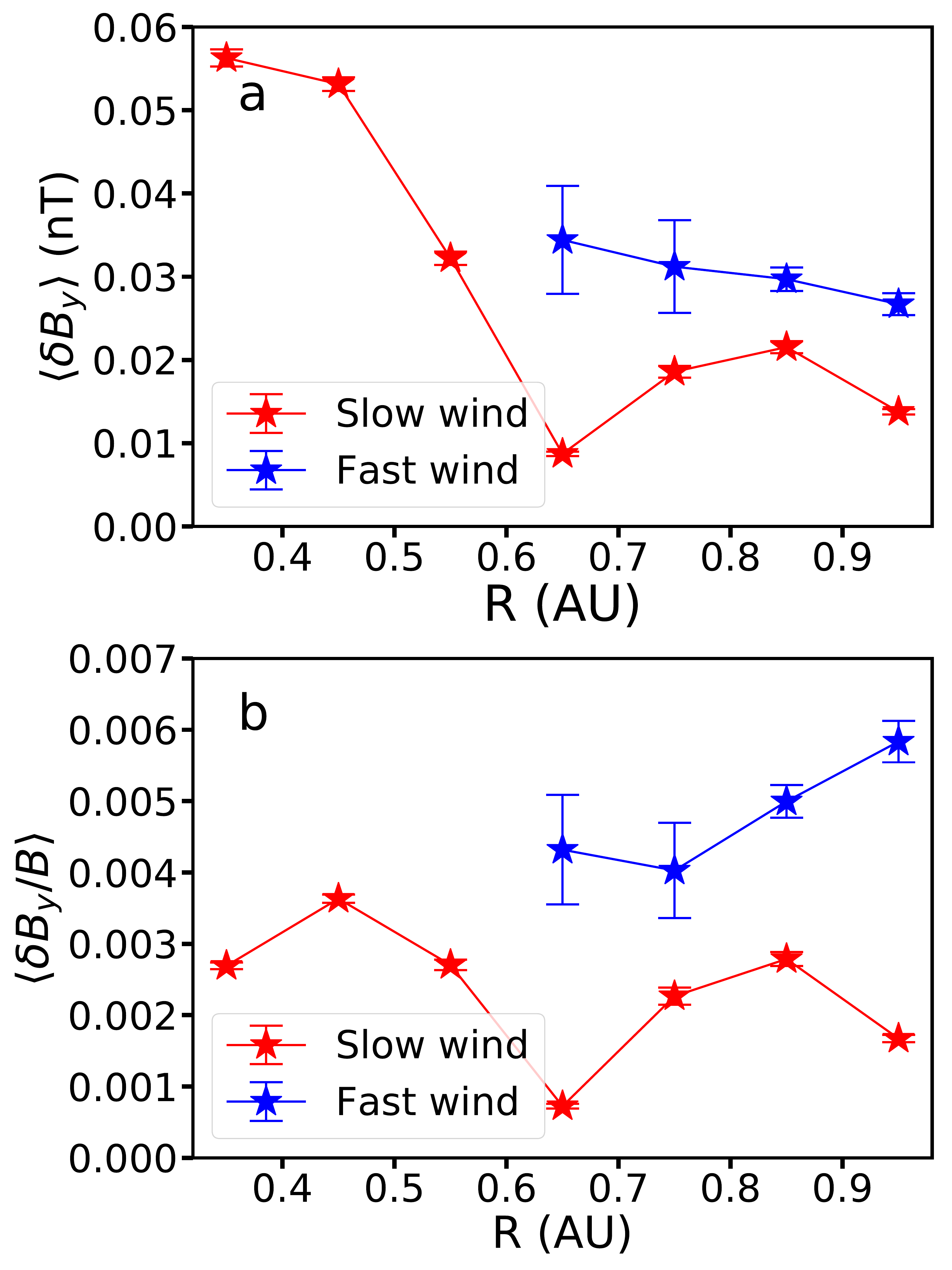}
%\decoRule
\caption{Panel (a) shows the amplitude of whistler waves and panel (b) shows the normalized amplitude of whistler waves in the slow and fast solar wind. Error bars show the standard error ($\frac{\sigma}{\sqrt{n}}$). }
\label{fig:whistler amplitude}
\end{figure}

Using the contribution from the $B_y$ spectral component we have calculated the approximate RMS amplitude of the fluctuations that are associated with whistler waves. 
The amplitude of whistler wave is estimated as in the Equation~(\ref{eq:amplitude}),
where the peak value ($PSD_y$) of the bump is multiplied with its respective frequency bandwidth ($\Delta f$), which gives us the mean square amplitude of the fluctuation. The square root of mean square amplitude can be interpreted as the amplitude of the fluctuation. Note that by this definition we do not have exact amplitudes of the wave, but rather the standard deviation of the associated fluctuating wavefield within the 8 seconds of the integration of the spectrum.
\begin{equation}\label{eq:amplitude}
    \delta B_y =\sqrt{{PSD_y}*\Delta f}
\end{equation}
We have not removed the background turbulence contribution as the amplitudes of the considered whistler waves are large enough that the former represents a minor fraction of the total amplitude ($< 10 \%$ ).

Figure \ref{fig:psd_whistlers} shows the radial variation of the mean power spectral densities corresponding to the peak of the spectral bumps for the case of the slow (red) and fast (blue) solar wind. Figure \ref{fig:whistler amplitude} (a) shows the radial variation of the mean whistler amplitudes separately for the case of slow (red) and fast (blue) solar wind.
We see that the whistler waves have higher amplitudes in fast than in slow wind and that there is no clear radial trend.

Figure \ref{fig:whistler amplitude} (b) shows the radial variation of the normalized whistler amplitudes ($\delta B_y/B$) for the  slow  and fast  solar wind. We have normalized the whistler amplitude values with the closest mean magnetic field values. Interestingly we observe that whistler waves have larger relative amplitudes in the fast wind than in the slow wind. We do not observe a clear trend in the radial evolution for both types of the wind. We infer that observed whistler waves may be considered to be in the linear regime (${\delta B}/{B} < 0.1$).

We have looked into different plasma parameters such as electron and proton density, electron and proton temperatures, electron core and halo anisotropies, magnetic field, electron and proton beta in order to identify a possible relation with the amplitudes of the observed whistlers. The core electron beta ($\beta_{ec}$) and halo electron beta ($\beta_{eh}$) shows the best possible correlation. For $\beta_{ec}$ and $\beta_{eh}$ we found to have a weak but statistically positive correlation of $\sim 0.40$ with the relative amplitudes.

\subsection{Estimation of phase velocity}

Using the cold plasma theory we have the dispersion relation of the right hand circularly polarized whistler waves \citep{Bellan2006},  
\begin{equation}
n^2=\frac{c^2k^2}{\omega^2}\approx 1+\frac{{\omega_{pe}^2 }/{\omega^2}}{(\frac{{\omega_{ce}}\cos\theta_{kB}}{{\omega}}-1)}.
\end{equation}
For parallel propagating waves with $\omega \ll \omega_{ce}$ we have $$ (\frac{{\omega_{ce}}\cos\theta_{kB}}{{\omega}}-1) \approx \frac{\omega_{ce}}{\omega}, $$
thus, the dispersion relation becomes:
\begin{equation}
{c^2k^2} \approx \omega^2 + \omega_{pe}^2(\omega/\omega_{ce}).
\end{equation}
For $\omega \ll \omega_{pe}$, we obtain
\begin{equation}
c^2k^2\approx \omega_{pe}^2(\omega/\omega_{ce}),
\end{equation}
Then, the phase velocity of the quasi parallel whistler is given by
\begin{equation}\label{eq phase velocity}
v_\phi=\frac{\omega}{k}=c\frac{\sqrt{\omega \omega_{ce}}}{\omega_{pe}}.
\end{equation}
The same equation~(\ref{eq phase velocity}) also holds for anti-parallel whistler waves.

Assuming that the spectral bumps correspond to the parallel whistlers and the observed frequency in the satellite frame $f$ is not much influenced by the Doppler shift and taking in to account low relative frequency of the observed wave signatures ($f\simeq 0.1 f_{ce}$), we can apply expression (\ref{eq phase velocity}) to estimate $v_\phi$ for our data set.

\begin{figure}
\centering
\includegraphics[width=0.8\linewidth]{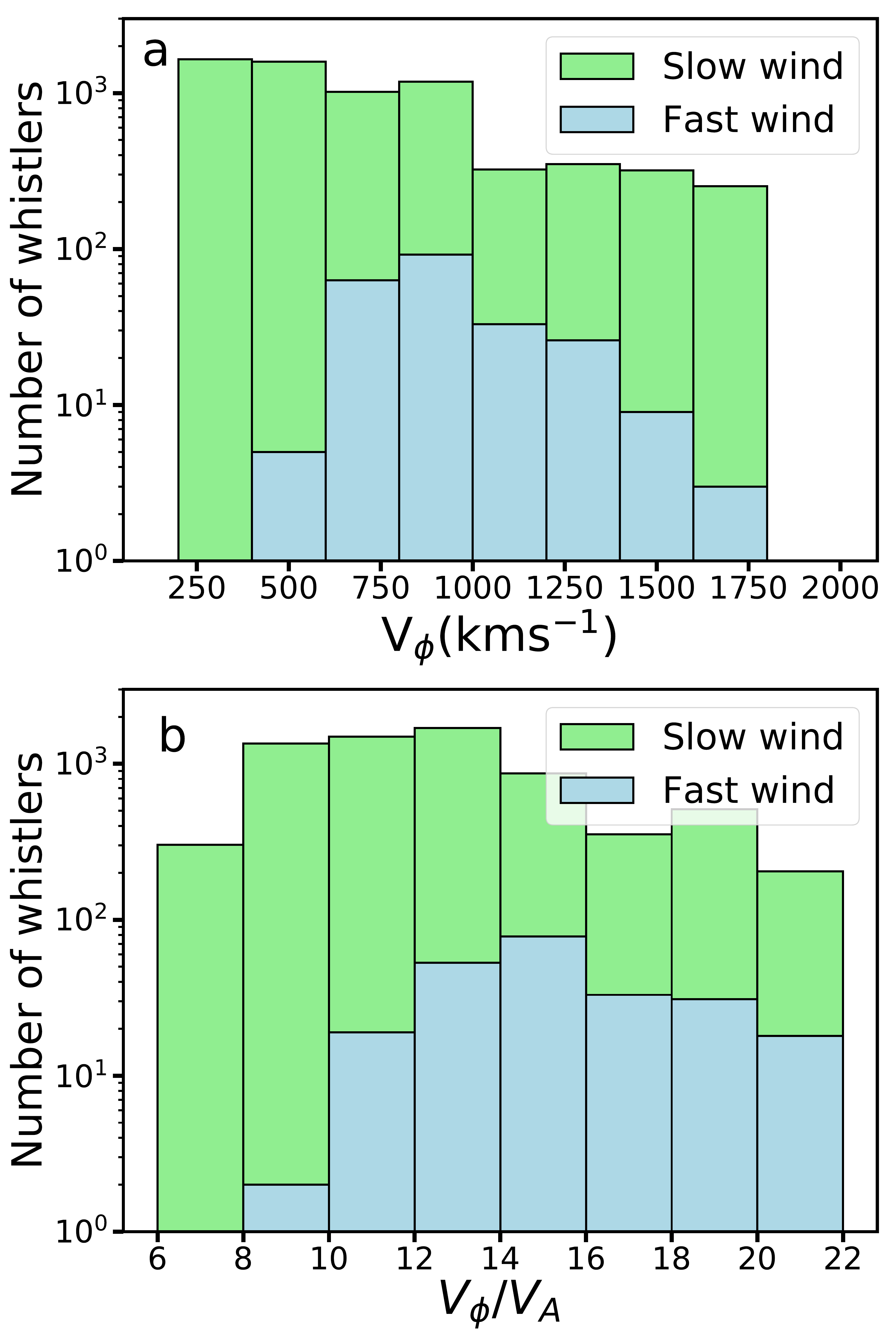} 
%\decoRule
\caption{Histograms of the phase velocity and normalized phase velocity of all the observed whistler waves. Green corresponds to the slow wind and blue to the fast wind. In panel (a) we show the phase velocity and in panel (b) we show the phase velocity normalised to their corresponding Alfv\'en speed ($V_A$).}
\label{fig:phase velocity}
\end{figure}

Figure \ref{fig:phase velocity} (a) shows the distribution of phase velocities ($V_\phi$) determined from the central frequencies of the spectral bumps in the 
fast wind (blue) and in the slow wind (green). We observe that phase velocities of the slow and fast solar wind whistlers are in the same range $2\cdot 10^2 <V_\phi<2\cdot 10^3$~km/s with the median around $\simeq 10^3$~km/s. Most of the whistler waves have high phase velocities with respect to $V_{sw}$ (not shown) and especially with respect to the Alfv\'en speed, see Figure \ref{fig:phase velocity} (b) where one observes $9 < V_\phi/V_A < 21 $ in the fast wind and $ 6 < V_\phi/V_A < 21 $ in the slow wind. We have also looked into how the phase velocity of whistlers is compared to the electron thermal velocity ($V_{th,e}$), we found that $ 0.5 < V_\phi/V_{th,e} < 0.8 $ (not shown).

\subsection{Electron anisotropy and heat flux  corresponding to the observed whistler waves}

\begin{figure}
\centering
\includegraphics[width=1.0\linewidth]{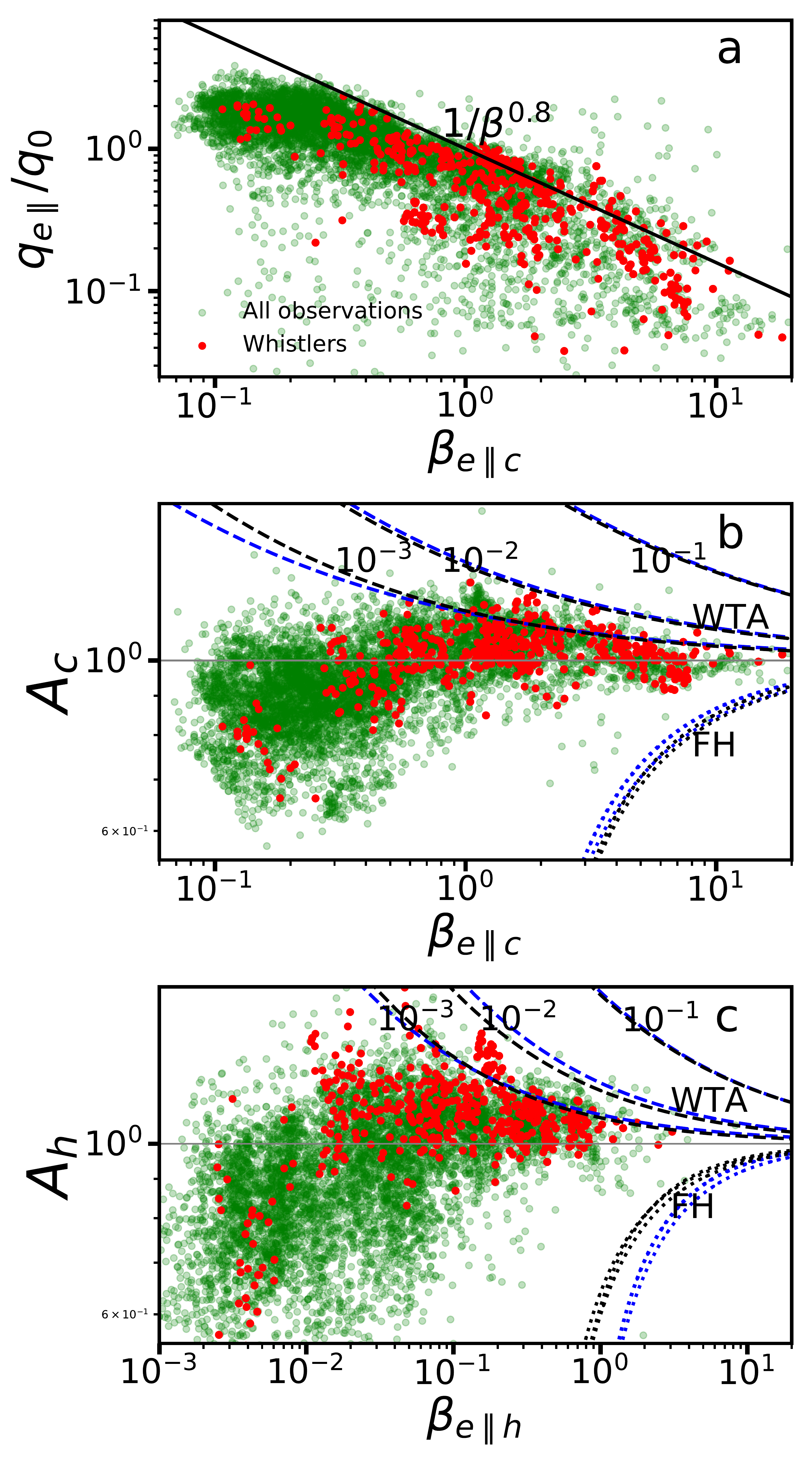}
%\decoRule
\caption{Normalized heat flux and core and halo temperature anisotropy during the period of spectra analyzed (12-12-1974 to 20-09-1975). Red dots represent the electron moments corresponding to the whistlers and the green dots correspond to all the electron moments measured during the availability of SCM data. In panel (a), we show the normalized heat flux ($q_{e\parallel}/q_0$) as a function of electron core parallel beta ($\beta_{e\parallel c}$), the blue line represents the heat flux instability threshold $\frac{1}{\beta_{c\parallel}^{0.8}}$ ($\frac{\gamma}{\omega_{ci}}=10^{-2}$), given by  \citep{Gary1999heatflux}. In panels (b) and (c), we show $A_c$ and $A_h$ as a function of $\beta_{e\parallel c}$ and $\beta_{e\parallel h}$ respectively. The maximum growth rate curves ($\frac{\gamma}{\omega_{ce}}=10^{-3}, 10^{-2},10^{-1}$) for the core and halo WTA instability are dashed; for the FH instability they are dotted and are taken from the work of \citet{Lazar2018}. The blue curve corresponds to $\kappa =3$ and the black curve to $\kappa =8$. } 
\label{fig:heat flux whistler}
\end{figure}

Studies such as \citet{Lacombe2014,Stansby2016,Tong2019} have suggested that heat flux instability might be acting when the whistlers are observed. Studies by \citet{Gary1999,Wilson2013} have shown that $A_{h}>1$ when whistler waves are observed. Recently \citet{Tong2019} have shown that if $A_{h}$ approaches 1 and higher, the plasma becomes unstable for the heat flux instability. The reason for $A_{h}$ significance will be discussed in detail in the coming sections. 

The electron moments on \textit{Helios}  are not always available: there are large gaps.
To increase the statistics we have considered the closest electron moment values to the observed whistlers. We kept the constraint of maximum 10~minutes from the nearest observed whistler. 
In Figure~\ref{fig:heat flux whistler} we present (a) the normalized electron heat flux $q_{e\parallel}/q_0$ as a function of the parallel beta of the core of the electron distribution function $\beta_{e\|c}$; (b) the electron temperature anisotropy of the core $A_{c}$ as a function of $\beta_{e\|c}$  and (c) the electron anisotorpy of the halo electrons $A_{h}$ as a function of $\beta_{e\|h}$. Green dots correspond to all the electron moments measured during the availability of SCM data; the subset of events for which whistler waves are observed is shown in red.

From Figure~\ref{fig:heat flux whistler}~(a) we infer that whistler heat flux instability (WHF) is probably at work as most of the data follow the marginal stability threshold of $\frac{\gamma}{\omega_{ci}}=10^{-2}$ \citep{Gary1999heatflux}). The $q_{e\parallel}/q_0$ values corresponding to the whistlers are observed close to the threshold. However, some values are below the threshold which suggests that the WHF instability might not be the sole cause of the whistlers. Other than $q_{e\parallel}/q_0$ values $A_h$ also has an important role as shown by \citet{Tong2019} that $A_h$ close and higher than one may significantly affect the onset of the WHF instability.  
Indeed, for our data set, we find that in 88$\%$ of spectra with signatures of whistlers, $A_h >1$. This ascertains the importance of $A_h$ for the whistler wave observations and for their generation through WHF.

Figure \ref{fig:heat flux whistler} (b) and (c) show $A_c$ and $A_h$ as a function of $\beta_{e\parallel c}$ and $\beta_{e\parallel h}$ respectively. We can infer that the core and halo whistler anisotropy instability is probably at work as most of the data are well constrained 
along the instability threshold contour $\frac{\gamma}{\omega_{ce}}=10^{-2}$. The maximum growth rate contours for the whistler temperature anisotropy (WTA) and firehose instability (FH) for the different cases of Kappa ($\kappa=3,8$) are taken from the work of \citet{Lazar2018}, where the authors have considered a bi-Maxwellian core and a bi-Kappa halo of the electron distribution function.
All the whistler points are closer to the whistler anisotropy instability threshold than to the firehose instability threshold. 

The large number of events located in the vicinity of the WTA threshold supports the role of the core and halo WTA instability in generating whistlers. We notice that when whistlers are observed, for most of the cases we find $A_c >1.0$ and $A_h >1.0$. We found that nearly 75$\%$ of the $A_c$ values corresponding to whistlers satisfied $A_c >1$ and as mentioned before 88$\%$ of the $A_h$ values satisfied $A_h>1$. We can also observe that the higher the $\beta$, the lower the instability thresholds, therefore the higher the probability of the instability to develop for high $\beta$ cases, as in the slow solar wind.

To summarize, Figure \ref{fig:heat flux whistler} suggests that the WHF and WTA instabilities are acting and that the three parameters $q_{e\parallel}/q_0$, $A_c$ and $A_h$ could be responsible for the observation of whistler wave signatures in the \textit{Helios} SCM spectra. 
However, due to the limitations on the data availability we cannot clearly state which type of instability is acting at which time and for how long, whether is it WHF or core WTA or halo WTA instability.

Previously, to have larger statistics we have considered the closest available electron moment values. Now, let us consider the electron properties that correspond to the observed whistler wave signatures. We first look for time intervals in which we observe at least 10 consecutive 8-sec spectra with whistlers, i.e. time intervals that are at least 80 s long. Then, we identify the electron moments which are measured in those whistler intervals. By this method we ensure that moments measured are always during the presence of whistler waves, we found only 45 such intervals. Their $q_{e\parallel}/q_0$ values are spread below and above the heat flux instability threshold. However, for all the observed whistler waves, we observe that $A_h >1.01$ in agreement with the studies of \citet{Gary1999,Wilson2013}. Our observations also agree with the work of \citet{Tong2019}, who showed the importance of the halo anisotropy ($A_h$) for the onset of heat flux instability.

Even though we do not have exact values of $q_{e\parallel}/q_0$, $A_h$ and $A_c$ for all the observed spectra with the signatures of whistler waves, we can have a general idea of the conditions around when the whistler waves appear, i.e. usually in the slow wind. $A_h$ and $A_c$ values are relatively higher in the slow solar wind when compared to the adjacent fast solar wind, while $q_{e\parallel}/q_0$ values are lower. A glimpse of this transformation can be seen from an example interval shown in Figure \ref{fig:plasma parameters}.

\begin{figure}
\centering
\includegraphics[width=0.9\linewidth]{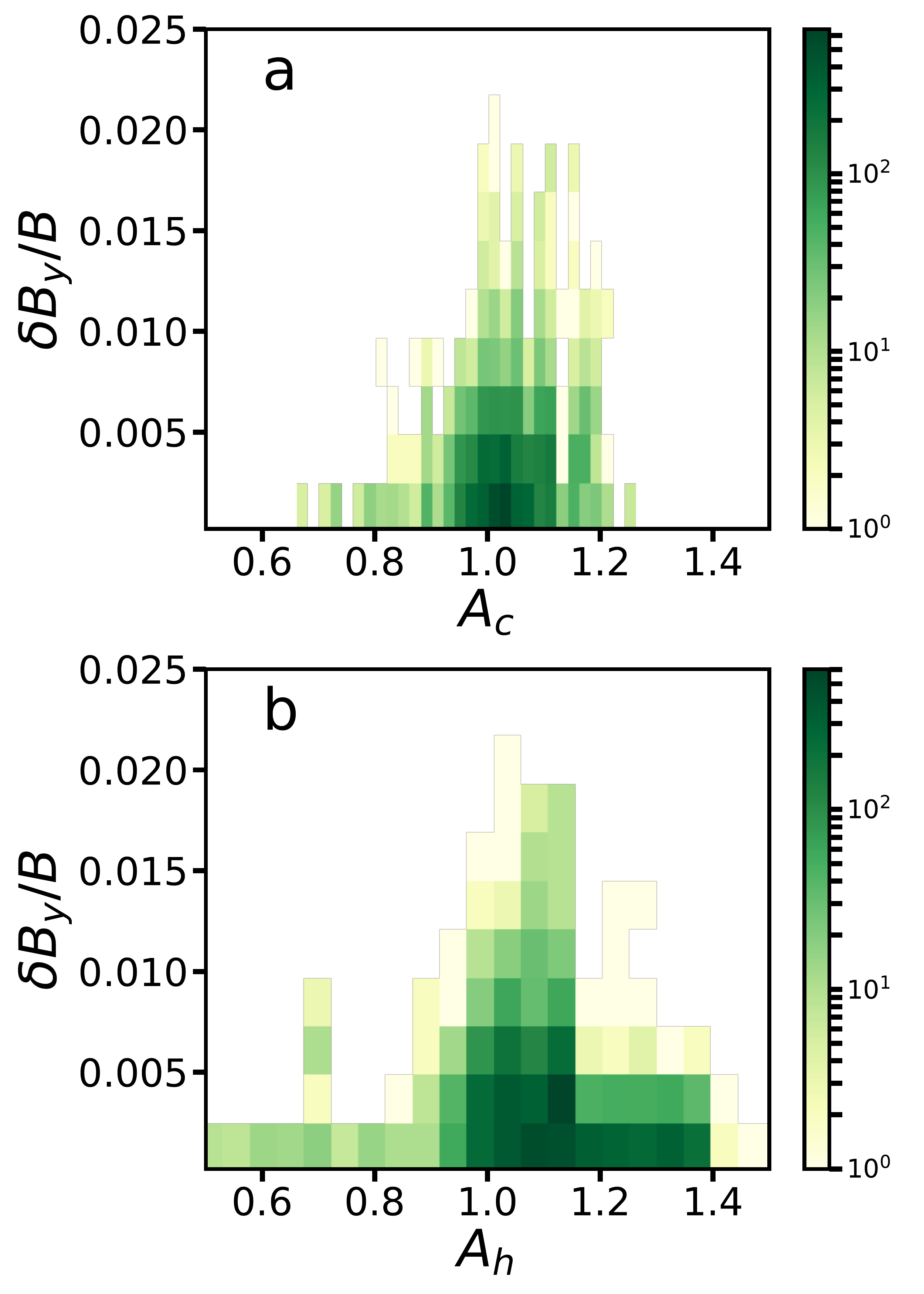}
%\decoRule
\caption{2d histogram of normalized amplitude as a function of electron core and halo anisotropies. In panel (a), we show the normalized amplitude variation as a function of $A_c$ and in panel (b), we present the normalized amplitude variation as a function of $A_h$.    }
\label{fig:Anisotropy}
\end{figure}

We have also looked into whether relative amplitudes have any relation to the $A_h$ and $A_c$ values.  From Figure \ref{fig:Anisotropy} (a) $\&$ (b) we can infer that 
(i) almost all the whistlers are observed when, $0.9 \le A_c \le 1.2$ and $0.9\le A_h \le 1.4$ and (ii) the relative amplitudes $\delta B_y/B$ are not dependent on the value of $A_c$ or $A_h$.

\section{Discussions}
\label{sec:Discussions}

\subsection{Why are whistler waves predominantly observed in the slow solar wind ?}

Different effects can explain the prevalence of whistler waves in the slow solar wind. These can be observational (whistler are present but cannot be observed), or physical (conditions are not met for generating whistlers, or existing whistlers are damped). Let us first consider the observational ones. 

One of the major differences between fast and slow winds is the higher relative fluctuation level of the magnetic field in the former, which may therefore limit the visibility of whistler waves, as noticed by \citet{Lacombe2014}. We performed tests to quantify the impact of the fluctuation level, see Appendix \ref{sec:visibility}. While this effect plays a role it is not sufficient to explain the much smaller number occurrence probability of whistlers in the fast solar wind. Another effect could be the enhanced Doppler shift in the fast wind. We also quantified this effect, which can be ruled out, see Appendix \ref{sec:visibility}. What is then left over are the conditions for whistler waves to be generated in the solar wind.

\subsection{Presence of whistler signatures and the conditions of whistler generation }

The two major sources of whistler wave generation in the solar wind are the whistler heat flux (WHF) and whistler temperature anisotropy (WTA) instabilities \citep{Garry1994}.
Observational studies at 1~AU show that whistlers can be generated by the WHF instability \citep{Lacombe2014,Tong2019}. We have already shown in Figure \ref{fig:heat flux whistler} (a) that the WHF instability conditions are indeed met for some whistlers in the inner heliosphere. Figures \ref{fig:heat flux whistler} (b) and (c) show that the WTA instability conditions are also met for core and halo electron populations. From this we conclude that the ratios $q_{e\parallel}/q_0$, $A_h$ and $A_c$ are important parameters for understanding the presence of whistler waves in the inner heliosphere. In addition, from the studies of \citet{Gary1977,Gary1999,Wilson2013,Tong2019} we understand that the ratio $\frac{T_{e\perp h}}{T_{e\parallel h}}$ is another important parameter for the onset of the WHF instability.

%%%%%%%%%%%%%%%%%%%%%%%%%%%%%%%%%%%%%
%%%%%%%%%%%%%%%%%%%%%%%%%%%%%%%%%%%
\citet{Gary1977}, assuming a two bi-maxwellian velocity distribution function approximation, derived the dispersion relation for the whistler waves, the growth rate is given as
\begin{equation}\label{dispersion gary}
\frac{\gamma}{\Omega_i} \propto \{(\textbf{k} \cdot \textbf{v}_{0H} -\omega_R)\frac{T_{e\perp h}}{T_{e\parallel h}}+\mid\Omega_e\mid(\frac{T_{e\perp h}}{T_{e\parallel h}}-1)\},   
\end{equation}
where $\Omega_e$ is the electron cyclotron frequency, $\Omega_i$ is the ion cyclotron frequency, $\textbf{v}_{0H}$ is the drift velocity for halo electrons, $\omega_R$ is the real frequency.

In the Equation \ref{dispersion gary}, the first term on the right hand side is the contribution of heat flux due to the drift of the electrons and the second term the contribution of the anisotropy to the growth of the instability. Studies by \citet{Gary1977} have suggested that when the halo is nearly isotropic, i.e. when $A_h\simeq 1$, and the heat flux contribution term is positive, the WHF instability is driven. An example of this has been recently reported by \citet{Tong2019} using the \textit{ARTEMIS} data. \citet{Gary1977} have also suggested that when $A_h>1$, the temperature anisotropy can strongly drive the WTA instability.

We have shown that some of the whistler spectra correspond well to the WHF instability threshold (see Figure \ref{fig:heat flux whistler} (a)), however, a majority of them lie below it. In contrast more than 88$\%$ of events with the whistlers have $A_h>1$ and may contribute positively to the growth of the wave. One should note that whistlers are not always observed when $A_h>1$ (see Figure \ref{fig:heat flux whistler} (c)). This can be due to several reasons, one of them being that the halo anisotropy alone cannot act as a source for the generation of whistler waves. 

Similarly, we have not always observed signatures of whistler waves even if the ratio $q_{e\parallel}/q_0$ is above the whistler heat flux instability threshold (see Figure \ref{fig:heat flux whistler} (a)). A similar observation was made by \citet{Tong2019, Tong2019stasticalstudy}. This leads to the conclusion that the heat flux itself might not always be able to generate whistler waves and a more complicated interplay between the heat flux and anisotropy is required.

\begin{figure}
\centering
\includegraphics[width=0.9\linewidth]{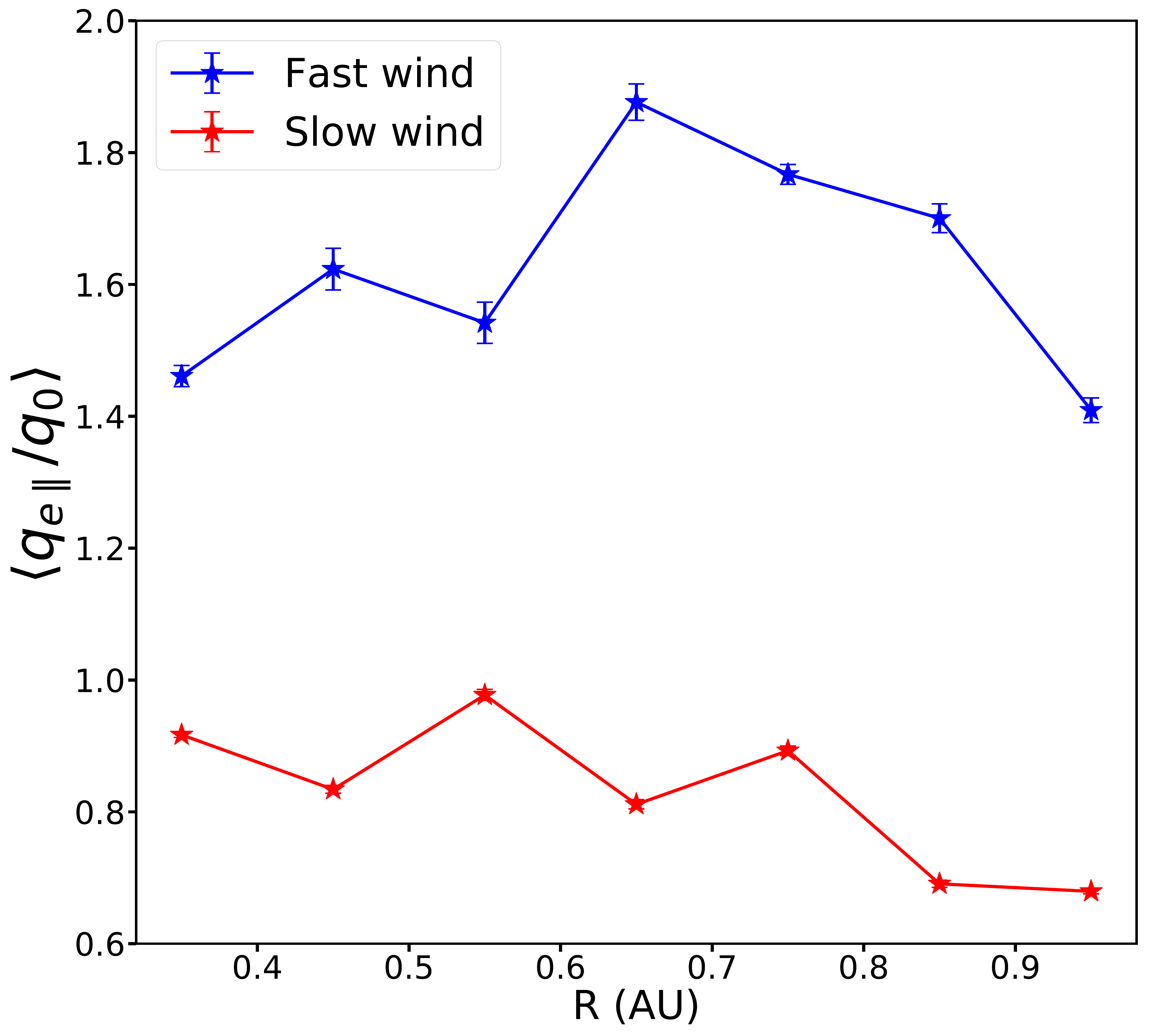}
%\decoRule
\caption{Mean of $q_{e\parallel}/q_0$ values for the slow (red) and fast (blue) solar wind as a function of distance from the Sun for the whole \textit{Helios} mission. The error bars here show the standard error ($\frac{\sigma}{\sqrt{n}}$) }
\label{fig:normalizedheatflux}
\end{figure}

Figure \ref{fig:normalizedheatflux} shows the radial evolution of the normalized heat flux ($q_{e\parallel}/q_0$), calculated by \citet{Stverak2015} for the period of 01-01-1975 to 10-05-1979, separately for the slow and the fast solar wind. Interestingly, we did not observe any correlation to the observed properties in Figure \ref{fig:percentage whistlers velocitySCM} and \ref{fig:whistler percentage}. Contrarily, we have observed that $q_{e\parallel}/q_0$ values are lower in the slow wind as compared to the fast wind. Furthermore as a function of radial distance $q_{e\parallel}/q_0$ values did not show any clear trends in the fast wind, whereas in the slow wind we observe a slight decrease in $q_{e\parallel}/q_0$ with $R$. These observations might be related to the impact of the whistler waves on the dissipation of the heat flux through the scattering of strahl electrons \citep{Stverak2015}.

\begin{figure}
\centering
\includegraphics[width=1.0\linewidth]{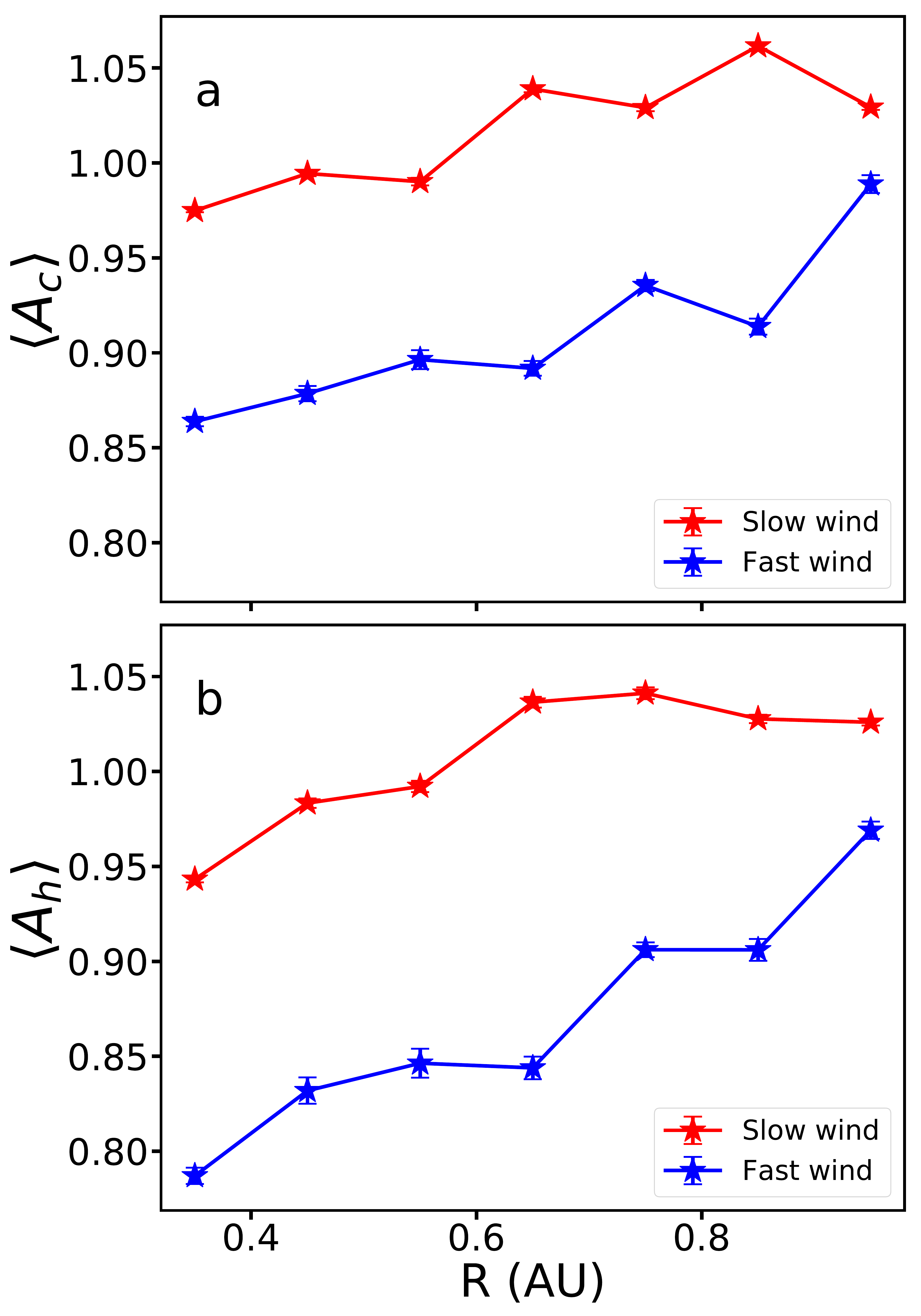}
%\decoRule
\caption{Illustration of electron anisotropy variations in the inner heliosphere for the slow (red) and fast (blue) solar wind for the whole \textit{Helios} mission. We present in panel (a) the mean $A_c$ ($\frac{T_{e\perp c}}{T_{e\parallel c}}$) and in panel (b) the mean $A_h$ ($\frac{T_{e\perp h}}{T_{e\parallel h}}$) values as a function of distance from the Sun. The error bars here show the standard error ($\frac{\sigma}{\sqrt{n}}$) }
\label{fig:heatflux}
\end{figure}

\begin{figure}
\centering
\includegraphics[width=1.0\linewidth]{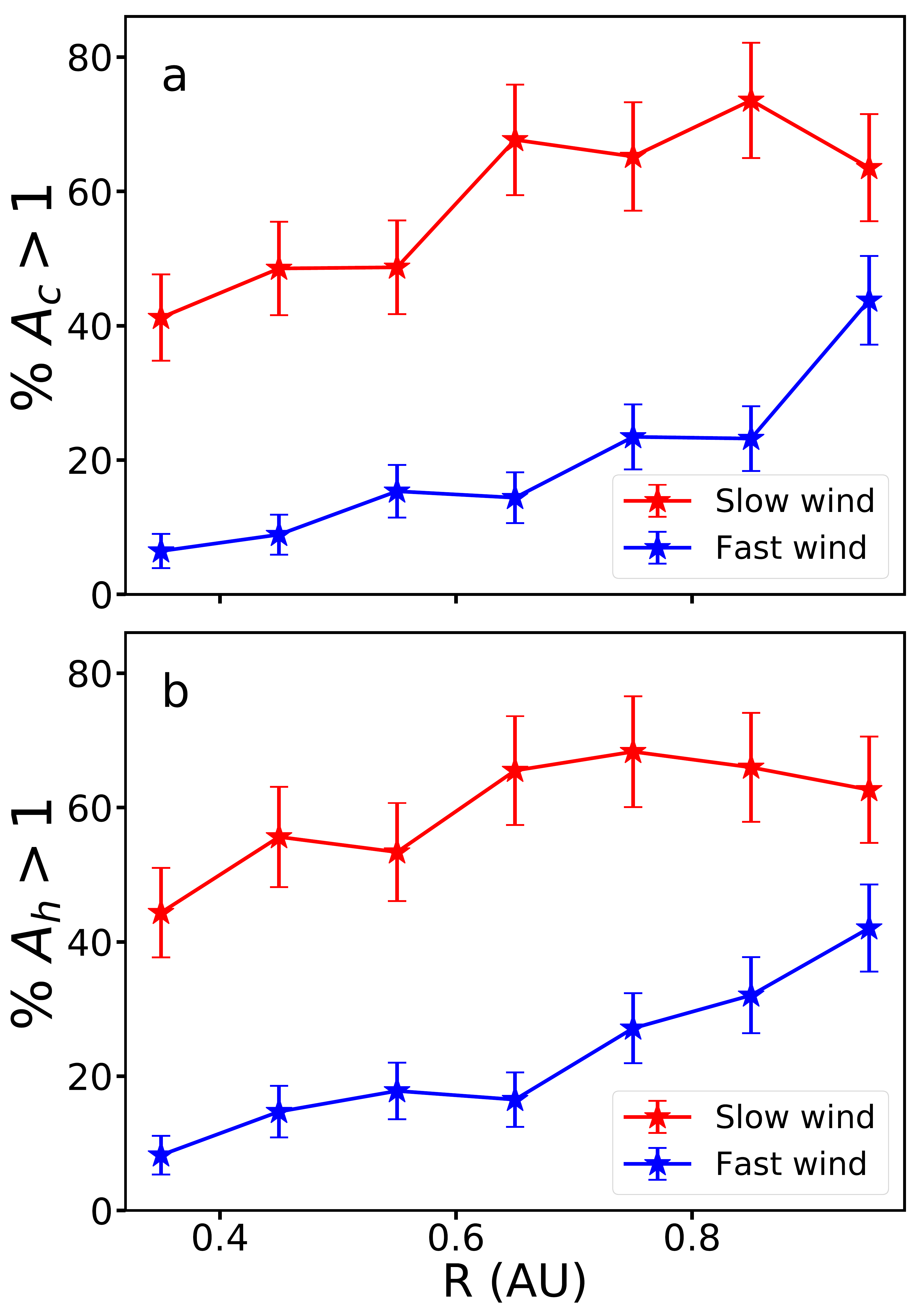}
%\decoRule
\caption{Illustration of electron anisotropy variations in the inner heliosphere for the slow (red) and fast (blue) solar wind for the whole \textit{Helios} mission. We show in panel (a) the percentage of $A_c$ $(\frac{T_{e\perp c}}{T_{e\parallel c}}) >1$ and in panel (b) the percentage of $A_h$ $(\frac{T_{e\perp h}}{T_{e\parallel h}}) >1$ for different velocity ranges. The error bars here show the standard error.}
\label{fig:aniostropies slow and fast wind}
\end{figure}

Figures \ref{fig:heatflux} and \ref{fig:aniostropies slow and fast wind} presents the radial evolution of the electron core and halo anisotropies, calculated by \citet{Stverak2009} for the period of 01-01-1975 to 10-05-1979. 
In Figure \ref{fig:heatflux} we show the variation of the mean electron core and halo anisotropies, while Figure \ref{fig:aniostropies slow and fast wind} shows the relative proportion of electron core and halo anisotropy values that are greater than one in the slow and fast winds as a function of radial distance.
We see that the values of $\langle A_h\rangle $, $ \langle A_c\rangle $ and also the percentage of electron core and halo anisotropy values that are greater than one ($ \% A_c >1  $,$ \% A_h >1  $) are higher in the slow solar wind throughout the inner heliosphere as compared to what we see in the fast solar wind. We also observe that as we move away from the Sun the value of $\langle A_h\rangle $, $ \langle A_c\rangle $  and also the $ \% A_c >1  $, $ \% A_h >1  $ increases both in the slow and fast solar wind. These observations point to two important conclusions. First, the conditions for the whistler generation are more favorable in the slow wind compared to the fast solar wind. Second, the conditions for whistler generation are improving as we move farther from the Sun.  

\begin{figure}
\centering
\includegraphics[width=0.9\linewidth]{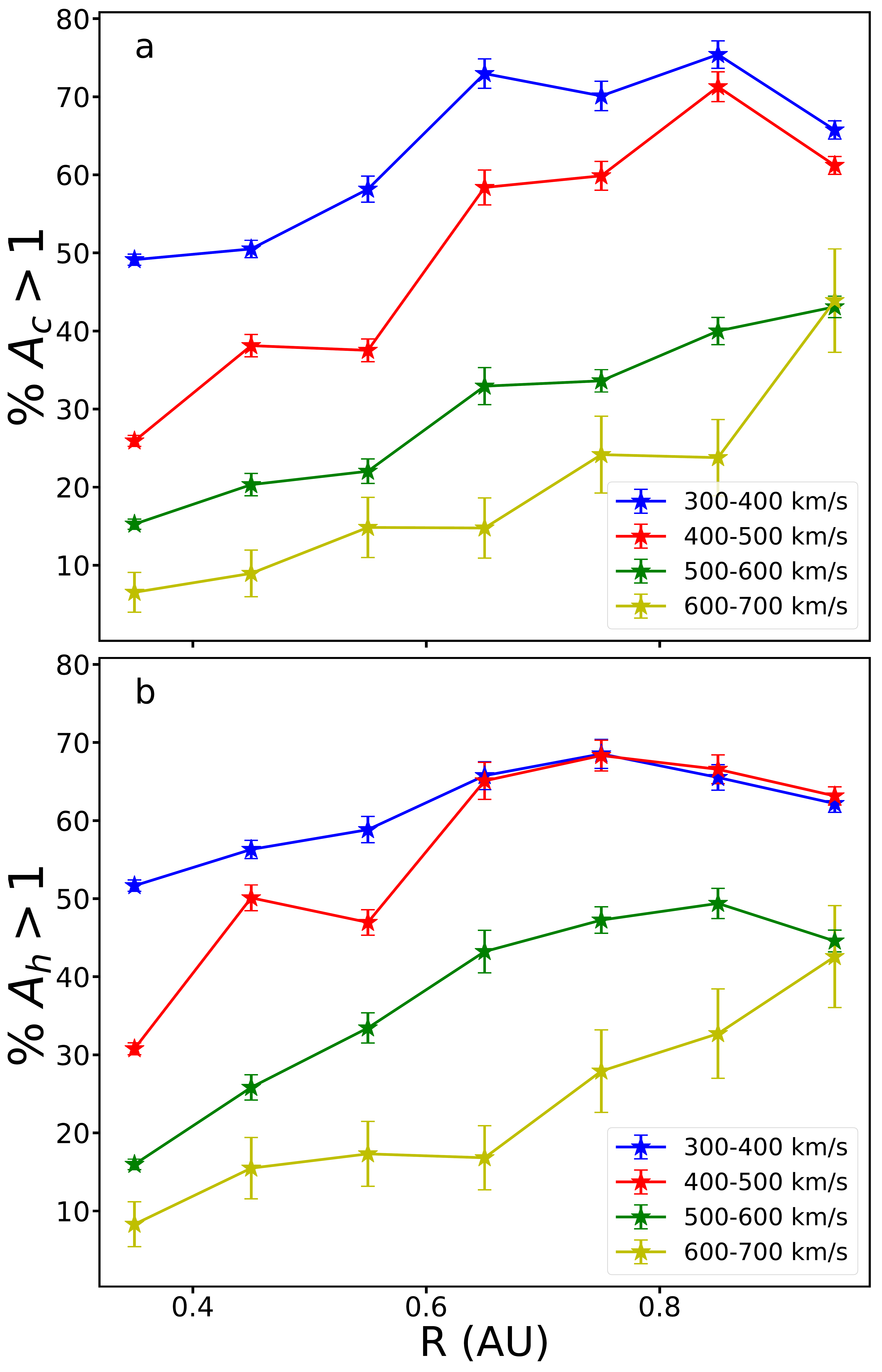}
%\decoRule
\caption{Illustration of electron anisotropy variations in the inner heliosphere for different types of wind for the whole \textit{Helios} mission. In panel (a) we show the percentage of $A_c$ $(\frac{T_{e\perp c}}{T_{e\parallel c}}) >1$ and in panel (b) we present $A_h$ $(\frac{T_{e\perp h}}{T_{e\parallel h}}) >1$ for different velocity ranges. The error bars here show the standard error.}
\label{fig:aniostropies velocities}
\end{figure}

In Figure \ref{fig:aniostropies velocities} we show the percentages of data with $ A_c >1  $ and $ A_h >1  $ in different wind types as functions of the radial distance.
Figure \ref{fig:aniostropies velocities} helps in explaining why there is a monotonic decrease in percentage of whistler waves with the wind velocity as we have observed in Figure \ref{fig:percentage whistlers velocitySCM}. There is a monotonic decrease of values $ A_h >1 $ and $ A_c >1 $ as the wind velocity increases and consequently we have a decrease in the percentage of whistler waves with velocity. We have also observed a similar monotonic decrease of $ \langle A_h \rangle$ and $ \langle A_c \rangle  $ with the increase of wind velocity (not shown). We would also like to mention that we have calculated how the distance from the $A_h$ and $A_c$ values to their respective WTA instability thresholds are varying from 0.3 to 1 AU. We found that as we move farther from the Sun, the $A_h$ and $A_c$ values are closer to the WTA instability thresholds, as the $\beta_{e\parallel c}$ and $\beta_{e\parallel h}$ are increasing with the distance and thresholds are lower for higher beta. Therefore, from this analysis too, we can understand that whistler generation conditions are improving as we move farther from the Sun.

\subsection{Reasons behind the observed $A_c$ and $A_h$ trends}

After discussing the observed signatures of whistler wave behaviors with the $A_c$ and $A_h$ values, some other questions can be raised. What can be the reason for the high $A_c$ and $A_h$ values in the slow wind as compared to the fast one ? What can be the reason for the increase in $A_c$ and $A_h$ as we move farther from the Sun?

\citet{stverak2008} showed that the value of $A_c$ increases with the electron collisional age.
Collisional age was reported to increase with radial distance and to be anti-correlated to the solar wind speed. This is in line with our observations for $A_c$.
Collisions are effective at low energies (core electrons), but cannot be effective at high energies, such as for halo electron population. We conjecture that the increase we observe in the $A_h(R)$ with $R$ and also the presence of whistler waves are closely related to the broadening (scattering) of the strahl part of the electron distribution. The basis for this idea is related to the observed properties of the strahl of the electron distribution.

Using the \textit{Helios} electron data, \citet{Laura2019} has shown that the strahl is in general broader in the slow than in the fast solar wind, and that the width in both cases increases with radial distance. Strahl width was reported to increase with radial distance by \citet{Hammond1996}, using data from \textit{Ulysses} mission, and by \citet{Graham2017}, using data from \textit{Cassini}.

Studies by \citet{Maksimovic2005} and by \citet{Stverak2009} show that the relative density of the halo and strahl vary oppositely over radial distance: while the strahl population is more pronounced closer to the Sun, the halo density increases farther from the Sun. These observations imply that the strahl is scattered into the halo over the radial distance.

The connection between the occurrence of whistler waves observed and the width of the strahl electrons - the strahl is broader and the occurrence of whistler waves is higher in the slow solar wind and farther from the Sun - suggests that the strahl might be scattered by the whistler waves. \citet{Kajdic2016} in his observational study at 1 AU had shown that the strahl is highly scattered in the presence of narrow-band whistler waves. In a different context, various numerical studies such as \citet{Vocks2005,Boldyrev2019,Tang2020} had shown that the interaction of electron VDF with the whistler wave turbulence could lead to the scattering of energetic strahl electrons and to the formation of an electron halo.

From the observed relation of whistler waves occurrence and the $A_h$ values, we conjecture that the observed changes in $A_h(R)$ might be related to the scattering of the strahl with the whistler waves.

\subsection{A feedback mechanism }
An important and still open question is how the momentum is transferred in this wave-particle interaction, to result in an increase of $A_h$.

\citet{Kajdic2016} observed a direct link between the scattering of the strahl and the presence of whistler waves. These whistler waves can interact with the strahl electrons through electron cyclotron resonance. In the following, we make a rough estimates of the energy ranges in which the whistler waves are able to interact resonantly.

The equation of electron cyclotron resonance is
\begin{equation}\label{eq2}
\omega -{\textbf{k}}\cdot\textbf{v}=\omega_{ce}\hspace*{2pt},
\end{equation}
\begin{equation}\label{eq2}
\omega -{{k}}{v}\cos\theta_{kv}=\omega_{ce}\hspace*{2pt},
\end{equation}
where, $\omega$ is the frequency of the wave, $k$ is the wave number of the whistler wave, 
$v$ is the velocity of the %strahl 
resonant electrons, 
$\theta_{kv}$ is the angle between the whistler wave wave vector and the velocity vector of the resonant electrons and $\omega_{ce}$ is the electron cyclotron frequency.

\begin{figure}
\centering
\includegraphics[width=1.0\linewidth]{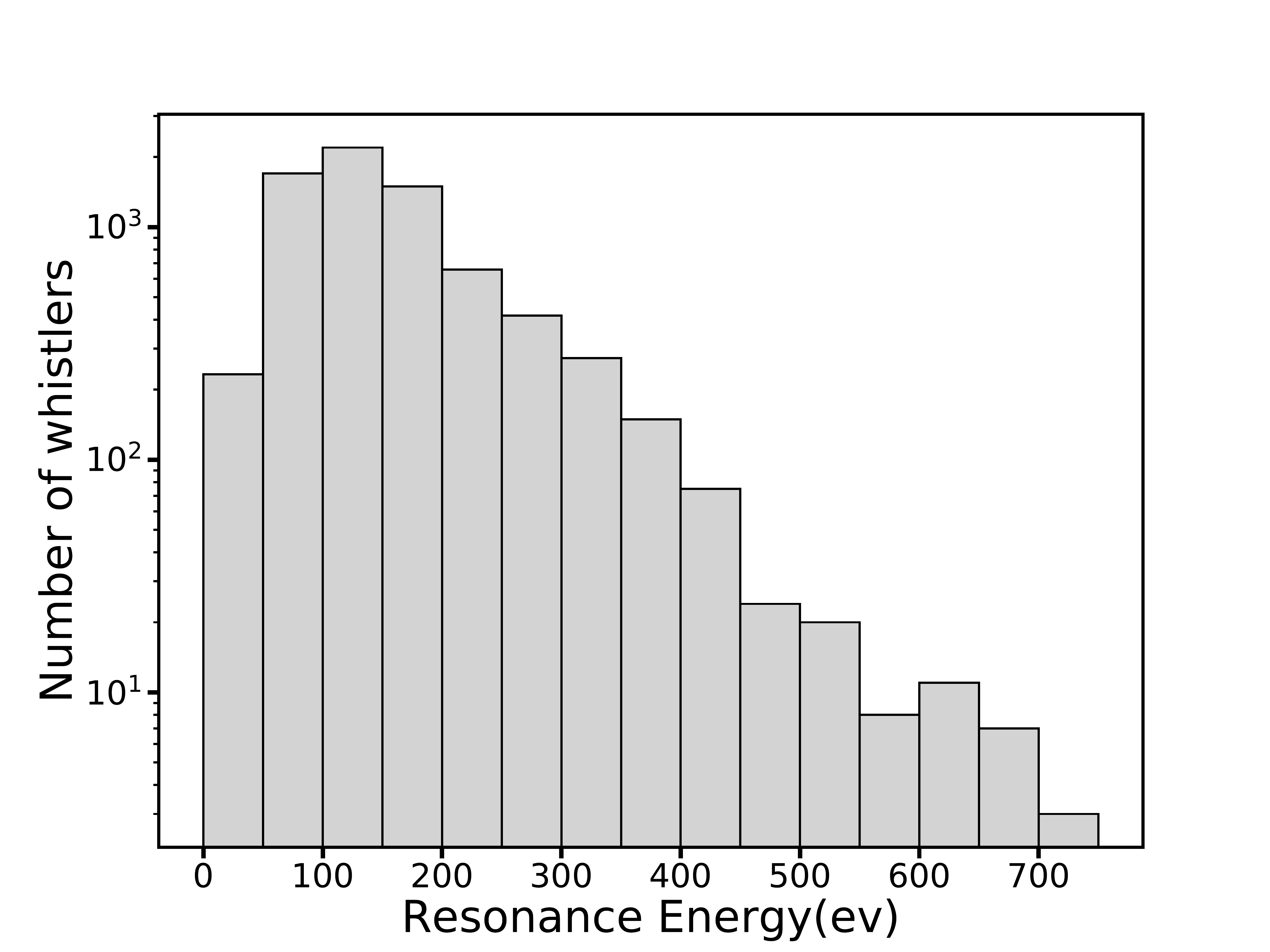}
%\decoRule
\caption{Resonant energy of the electrons interacting with the whistler waves (assuming that whistler waves travel opposite to electrons). }
\label{fig:Resonant Energy}
\end{figure}

Let us consider the case when resonant electrons are field aligned, as the strahl electrons \citep{Owens2017}. Most of the whistlers observed to date are quasi parallel or anti-parallel \citep{Lacombe2014,Stansby2016,Tong2019}. The condition for resonance is only satisfied when whistler waves travel opposite to the electrons as $\omega_{ce}$ is always greater than the frequency of the observed whistler waves. Therefore, we can understand that we have resonance when electrons travel in the opposite way to whistlers.
Assuming this condition, i.e. $\cos\theta_{kv}\approx -1$, we have
\begin{equation}
\omega +{k}{v}=\omega_{ce} ,
\end{equation}
\begin{equation}
\frac{{k}{v}}{\omega}=\frac{(\omega_{ce}-\omega)}{\omega},
\end{equation}
\begin{equation}\label{res_eq}
v={(\frac{\omega_{ce}}{\omega}-1)}{v_\phi}.
\end{equation}
Equation~(\ref{res_eq}) gives us the velocity of the electrons that can be resonant with the whistler waves (under the assumption of the opposite propagation between the wave and the electrons). 
The kinetic energy of these resonant electrons, 
 \begin{equation}\label{eq energy_res}
E=\frac{1}{2}m_ev^2,
\end{equation}
is shown in Figure~\ref{fig:Resonant Energy}. Most of the resonant electron energies are above 50 eV, which are the energies representative of the strahl and the halo populations. We conclude that the observed whistler waves might be able to scatter strahl and halo electrons through the electron cyclotron resonance. But how is the energy transferred in this type of interaction?

\citet{Veltri1993} have shown that when resonantly interacting whistler waves scatter the electrons, electron energy is transferred from the parallel to the perpendicular direction. Which could be understood as an increase of the temperature perpendicular to the magnetic field direction. On the other hand scattering the beam like population towards higher pitch-angles could be interpreted as scattering the strahl into the halo component.

We explain this with a feedback mechanism:
\begin{itemize}
    \item Whistler waves scatter the strahl population.
    \item Scattered strahl increases the value of $A_h$.
    \item Increases in the $A_h$ favors the creation of the whistler waves and on.
\end{itemize}

Therefore, we can hypothesize that whistler waves could scatter the strahl and create a self-sustaining mechanism that could explain the observed whistler properties.

However, the $A_h$ values cannot keep on increasing without a limit as they are bounded by whistler temperature anisotropy instability thresholds. This is the reason why we believe that there is a saturation of $A_h$ values in the slow wind farther from the Sun as shown in Figures \ref{fig:heatflux} and \ref{fig:aniostropies slow and fast wind}.

Recently \citet{Tang2020} has shown that whistler wave turbulence interaction with suprathermal electrons can lead to the formation of halo from strahl and could explain the observed opposite variation in the relative density of the halo and strahl with radial distance. In a future paper, we would like to verify our suggested feedback mechanism by modifying the kinetic transport model used in \citet{Tang2020}. We would like to treat halo and strahl as separate populations and study the interaction of intermittent narrow-band whistler waves with strahl and their influence on halo population.

Even though our explanation appears to be a reasonable one, there are some issues related to it. Our calculations are based on the assumption that $\cos\theta_{kv} \approx -1$ for the observed whistlers and resonant electrons. This might not always be true.  
Another case is that the whistler waves could be oblique. 
Recently, \citet{Vasko2019} using the linear stability analysis has shown that highly oblique whistler waves drive the pitch-angle scattering of strahl electrons, and in turn isotropize the halo and also suppress the heat flux. However, as we discuss in the introduction, from the previous observations in the solar wind we know that the oblique whistlers are usually associated with the large scale discontinuities \citep{Breneman2010J}. In the free solar wind, narrow-band whistlers are mostly quasi-parallel, e.g., \citep{Lacombe2014,Kajdic2016,Stansby2016}.

\section{Conclusion}
\label{sec:conclusion}
Our analysis of \textit{Helios} 1 data has revealed bursts of quasi-monochromatic waves in fluctuating magnetic field in the inner heliosphere, from 0.3 to 1 AU. These bursts are inferred from the power spectral density (PSD) of the magnetic field, for frequencies between 7 and 147 Hz. Although the absence of polarisation measurements prevents us from further characterizing these waves, based on the knowledge from other space missions (\textit{Cluster}, \textit{Stereo}, \textit{Artemis}, \textit{Geotail}) observations, the spectral bumps in \textit{Helios}/SCM measurements in the free solar wind are very likely due to the whistler waves. The radial dependence of the properties of these waves offer a unique opportunity for investigating the connection between their presence and the properties of the solar wind.

These whistler waves are predominantly observed in the slow solar wind: their probability of occurrence decreases with increase in $V_{sw}$.  
At the same time, the probability of occurrence of the signatures of whistlers increases with the radial distance from the Sun $R$. 
In the fast wind close to the Sun we do not observe any signatures of waves, they start to appear at $R>0.6$~AU.

These waves may be considered as being in a linear regime as their relative amplitude is generally smaller than 1\%. The amplitude is positively correlated with the plasma $\beta$ for the electron core and halo populations. For cases where simultaneous electron measurements are available, we find that the electron core anisotropy $A_c=\frac{T_{e\perp c}}{T_{e\parallel c}} > 1.01$. Our analysis suggests that both the whistler heat flux (WHF) and whistler temperature anisotropy (WTA) instabilities are at work. This is consistent with our observation that both the electron core anisotropy $A_c$ and electron halo anisotropy $A_h$ are bounded by the WTA instability threshold.

We show how the normalized heat flux was varying in the slow and fast solar wind with the radial distance. The normalized heat flux is found to be higher in the fast wind compared to the slow wind. We observe a decrease in heat flux with the radial distance. The signatures of whistler waves we have detected might explain the higher heat flux dissipation in the slow wind as compared to the fast wind and also the decrease of heat flux with the radial distance. 

The $\langle A_c \rangle$, $\langle A_h \rangle$ and the percentage of events for which $A_c>1$, $A_h>1$ are found to be higher in the slow solar wind as compared to the fast wind and their values increases as we move from 0.3 to 1 AU. The radial evolution and dominance in the slow wind of the core and halo anisotropies are similar to what is observed for the whistler occurrence.

%%%%%%%%%%%
%%%%%%%%%%%%

Based on these results we propose a connection between the observed whistler wave signatures and the observations of electron strahl. Indeed, the strahl is expected to be broader (i.e. more scattered) in the slow wind, in which whistler waves occur much more frequently. Conversely, the broadening of the strahl offers favorable conditions for whistler generation, which could then explain the larger presence of whistlers away from the Sun, and in the fast solar winds.

We proposed a possible mechanism for causing the $A_h$ ratio to vary in accordance with the occurrence of whistlers. Whistler waves scatter the strahl population, the scattered strahl increases the $A_h$ ratio which in turn favors the creation of whistler waves and so on. This may be considered as a feedback mechanism.

New observations by \textit{Parker Solar Probe} and \textit{Solar Orbiter} will provide a unique opportunity to verify our observed results with much better resolved magnetic and electron particle data. From our study, we speculate that \textit{Parker Solar Probe} and \textit{Solar Orbiter} will observe a relatively low occurrence of whistler waves than what we have observed at 0.3 AU as we go closer to the Sun in the slow wind. For the fast wind case, whistler waves might not be observed, even if they are observed they should be sparse. We also speculate that because of the saturation of the strahl, the likelihood of occurrence of whistler waves is likely to remain stable beyond 5 AU, both in the slow and fast solar winds.

\section{Appendix}

\subsection{Are visibility and high Doppler shift the sole reason for negligible fast wind whistlers ? }
\label{sec:visibility}

\begin{figure}[th]
\centering
\includegraphics[width=0.9\linewidth]{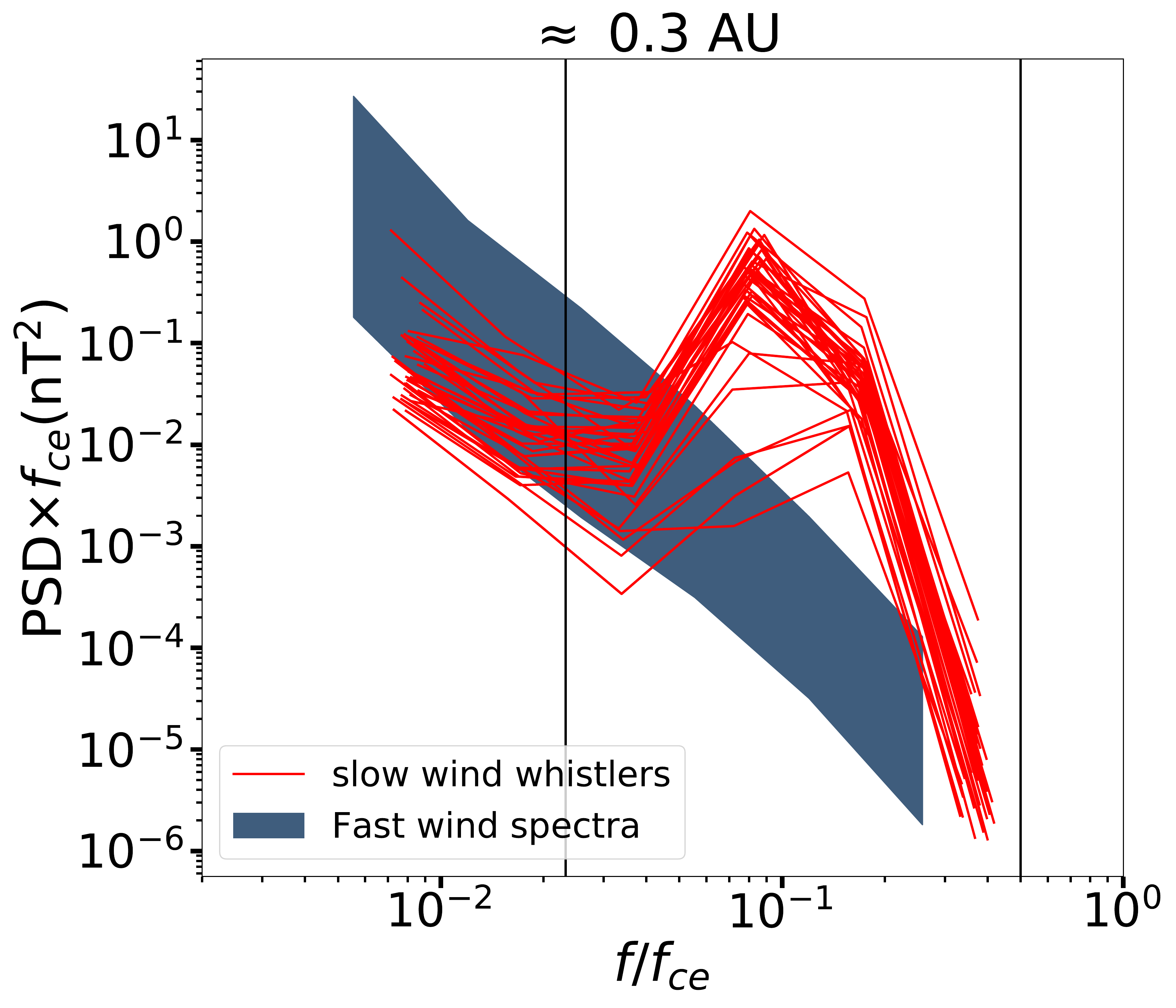}
%\decoRule
\caption{Spectra observed in the fast wind for 1 day (Day 73-74, 1975) and whistler spectra observed in slow wind for 1 day (Day 68.5-69.5, 1975). The vertical lines correspond to $f_{lh}$ and $0.5f_{ce} $. }
\label{fig:whistlers turbulence 0.3}
\end{figure}

\begin{figure}%
\centering
\subfigure(a){%
\label{fig:first}%
\includegraphics[width=0.9\linewidth]{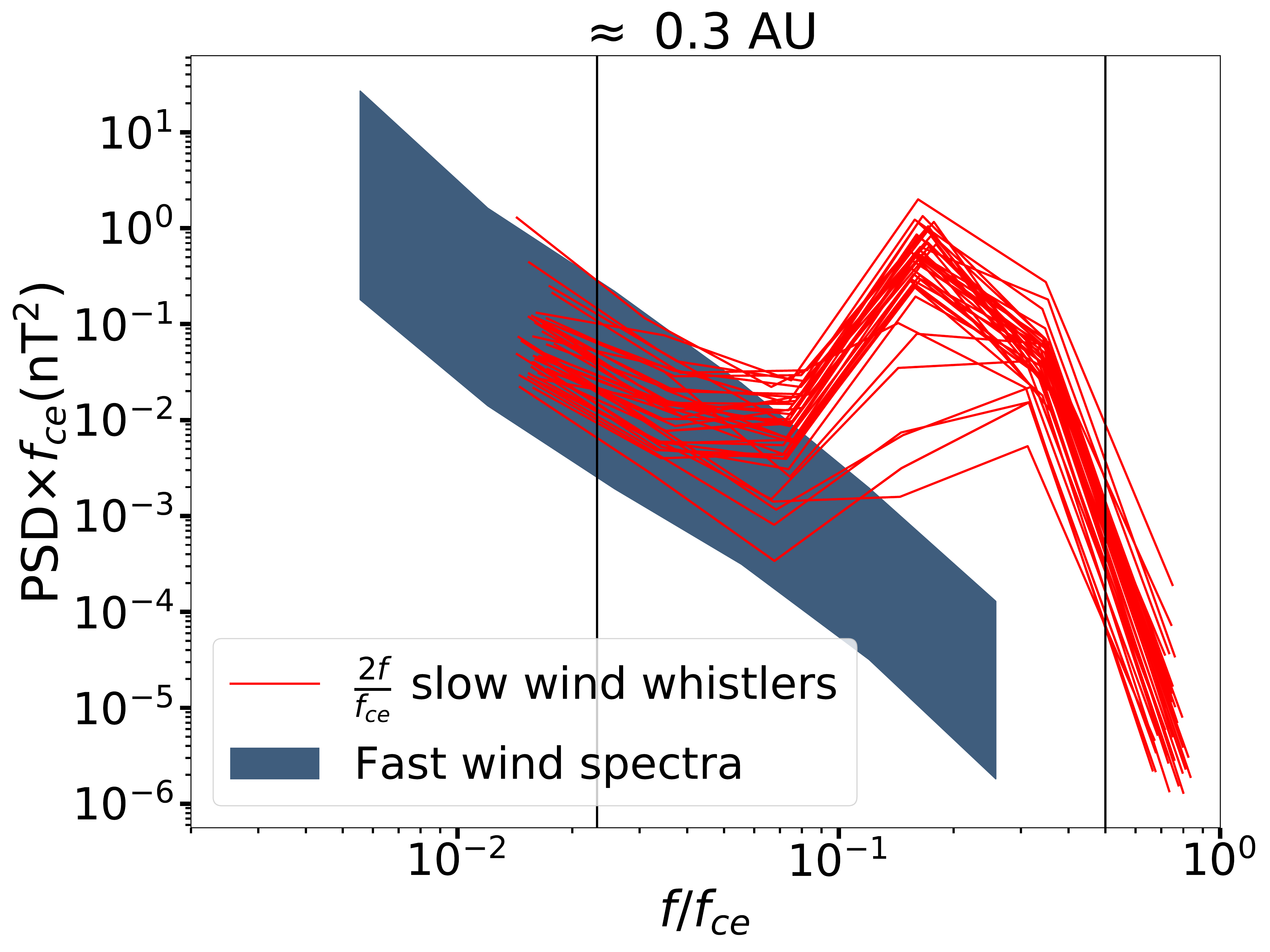}}%
\qquad
\subfigure(b){%
\label{fig:second}%
\includegraphics[width=0.9\linewidth]{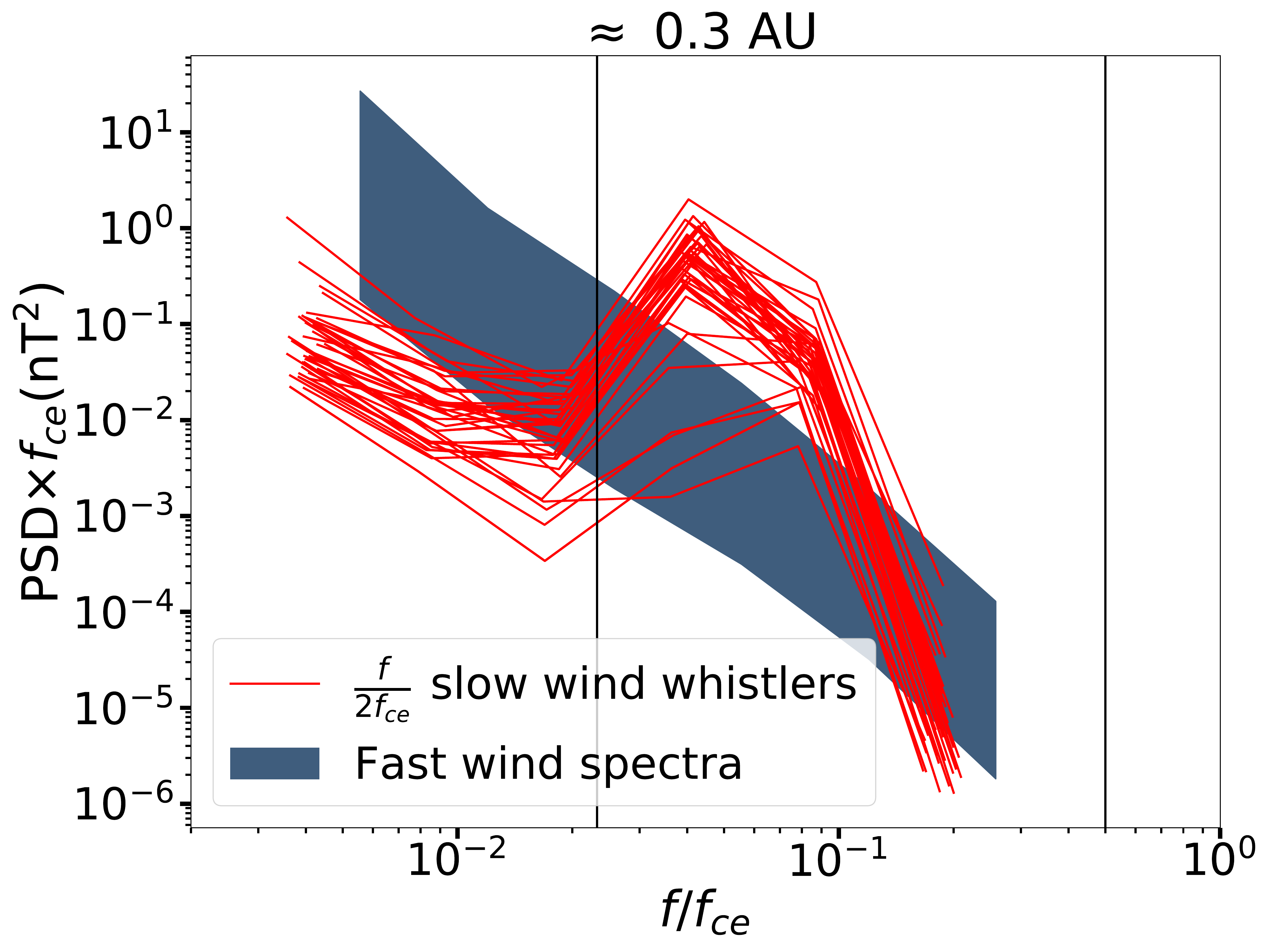}}%
\caption{Slow wind whistler waves extrapolated considering the Doppler effect in two extreme cases (a)  High frequency case  (b) Low frequency case. The vertical lines correspond to $f_{lh}$ and $0.5f_{ce} $. }
\label{fig:whistler turbulence doppler}
\end{figure}

To look into this issue, we have considered PSDs which are above 3 times the SCM noise from one day of slow wind and one day of fast wind at $\approx 0.3$ AU. We have separated the observed whistler waves in the slow wind and all the observed spectra in the fast wind. We show them together in Figure~\ref{fig:whistlers turbulence 0.3}, in which we plot the PSD as a function of $f/f_{ce}$, so that magnetic field differences in the slow and fast wind are taken into account.

From Figure \ref{fig:whistlers turbulence 0.3}, we get the first glimpse that, if intense whistler waves as in the slow wind are present in fast wind, they would be clearly visible as the turbulence level in the fast wind is not high enough to hide these types of waves. 

However, there could be another important issue of difference in the Doppler shift between the whistlers of slow and fast wind, as velocity of the fast wind is nearly twice that of the slow wind. To verify this, we have considered both the possible extreme cases, Doppler shift towards high frequency (2f) Figure \ref{fig:whistler turbulence doppler} (a) and Doppler shift towards low frequency (f/2) in Figure \ref{fig:whistler turbulence doppler} (b). From this Figure \ref{fig:whistler turbulence doppler} we can understand that if the whistler waves as shown here are present in the fast wind, the Doppler effect and turbulence level can be a reason but not the main issue to observe whistler waves. The high level of turbulence cannot hide these whistler waves.

\newpage

\section{Acknowledgement}
VKJ is supported by the French Space Agency (CNES) and R\'egion Centre-Val de Loire PhD grant (CNES N$^{0}$ 5100016274).  The authors would like to thank Catherine Lacombe and Lorenzo Matteini for their valuable suggestions and discussions. 
The \textit{Helios} data are available on the 
Helios Data Archive: \href{http://helios-data.ssl.berkeley.edu/}{http://helios-data.ssl.berkeley.edu/}. Plots for the article are made using Matplotlib \citep{Hunter2007}.

\bibliographystyle{yahapj}
\bibliography{references}

\begin{thebibliography}{}
\providecommand\natexlab[1]{#1}
\providecommand\JournalTitle[1]{#1}

\bibitem[{{Alexandrova} {et~al.}(2020){Alexandrova}, {Krishna Jagarlamudi},
  {Rossi}, {Maksimovic}, {Hellinger}, {Shprits}, \&
  {Mangeney}}]{Alexandrova2020}
{Alexandrova}, O., {Krishna Jagarlamudi}, V., {Rossi}, C., {et~al.} 2020,
  \JournalTitle{arXiv e-prints}, arXiv:2004.01102

\bibitem[{{Alexandrova} {et~al.}(2012){Alexandrova}, {Lacombe}, {Mangeney},
  {Grappin}, \& {Maksimovic}}]{Alexandrova2012}
{Alexandrova}, O., {Lacombe}, C., {Mangeney}, A., {Grappin}, R., \&
  {Maksimovic}, M. 2012,
  \href{http://dx.doi.org/10.1088/0004-637X/760/2/121}{\JournalTitle{The
  Astrophysical Journal}, 760, 121}

\bibitem[{{Artemyev} {et~al.}(2016){Artemyev}, {Agapitov}, {Mourenas},
  {Krasnoselskikh}, {Shastun}, \& {Mozer}}]{artemyev16}
{Artemyev}, A., {Agapitov}, O., {Mourenas}, D., {et~al.} 2016,
  \href{http://dx.doi.org/10.1007/s11214-016-0252-5}{\JournalTitle{Space
  Science Reviews}, 200, 261}

\bibitem[{{Beinroth} \& {Neubauer}(1981)}]{Beinroth1981}
{Beinroth}, H.~J., \& {Neubauer}, F.~M. 1981,
  \href{http://dx.doi.org/10.1029/JA086iA09p07755}{\JournalTitle{Journal of
  Geophysical Research}, 86, 7755}

\bibitem[{{Bellan}(2006)}]{Bellan2006}
{Bellan}, P.~M. 2006, {Fundamentals of Plasma Physics} (Cambridge,UK: Cambridge
  University Press)

\bibitem[{{Ber{\v{c}}i{\v{c}}} {et~al.}(2019){Ber{\v{c}}i{\v{c}}},
  {Maksimovi{\'c}}, {}, {Land i}, \& {Matteini}}]{Laura2019}
{Ber{\v{c}}i{\v{c}}}, L., {Maksimovi{\'c}}, {}, M., {Land i}, S., \&
  {Matteini}, L. 2019,
  \href{http://dx.doi.org/10.1093/mnras/stz1007}{\JournalTitle{\mnras}, 486,
  3404}

\bibitem[{{Boldyrev} \& {Horaites}(2019)}]{Boldyrev2019}
{Boldyrev}, S., \& {Horaites}, K. 2019,
  \href{http://dx.doi.org/10.1093/mnras/stz2378}{\JournalTitle{\mnras}, 489,
  3412}

\bibitem[{{Bothmer} \& {Schwenn}(1998)}]{Bothmer1998}
{Bothmer}, V., \& {Schwenn}, R. 1998,
  \href{http://dx.doi.org/10.1007/s00585-997-0001-x}{\JournalTitle{Annales
  Geophysicae}, 16, 1}

\bibitem[{{Breneman} {et~al.}(2010){Breneman}, {Cattell}, {Schreiner},
  {Kersten}, {Wilson}, {Kellogg}, {Goetz}, \& {Jian}}]{Breneman2010J}
{Breneman}, A., {Cattell}, C., {Schreiner}, S., {et~al.} 2010,
  \href{http://dx.doi.org/10.1029/2009JA014920}{\JournalTitle{Journal of
  Geophysical Research (Space Physics)}, 115, A08104}

\bibitem[{{de Lucas} {et~al.}(2011){de Lucas}, {Schwenn}, {dal Lago}, {Marsch},
  \& {Cl{\'u}a de Gonzalez}}]{lucas2011}
{de Lucas}, A., {Schwenn}, R., {dal Lago}, A., {Marsch}, E., \& {Cl{\'u}a de
  Gonzalez}, A.~L. 2011,
  \href{http://dx.doi.org/10.1016/j.jastp.2010.12.011}{\JournalTitle{Journal of
  Atmospheric and Solar-Terrestrial Physics}, 73, 1281}

\bibitem[{{Dehmel} {et~al.}(1975){Dehmel}, {Neubauer}, {Lukoschus},
  {Wawretzko}, \& {Lammers}}]{Dehmel1975}
{Dehmel}, G., {Neubauer}, F.~M., {Lukoschus}, D., {Wawretzko}, J., \&
  {Lammers}, E. 1975, \JournalTitle{Raumfahrtforschung}, 19, 241

\bibitem[{{Gary}(1993)}]{Gary1993}
{Gary}, S.~P. 1993, {Theory of Space Plasma Microinstabilities} (Cambridge,UK:
  Cambridge University Press)

\bibitem[{{Gary} \& {Feldman}(1977)}]{Gary1977}
{Gary}, S.~P., \& {Feldman}, W.~C. 1977,
  \href{http://dx.doi.org/10.1029/JA082i007p01087}{\JournalTitle{Journal of
  Geophysical Research}, 82, 1087}

\bibitem[{{Gary} {et~al.}(1994){Gary}, {Scime}, {Phillips}, \&
  {Feldman}}]{Garry1994}
{Gary}, S.~P., {Scime}, E.~E., {Phillips}, J.~L., \& {Feldman}, W.~C. 1994,
  \href{http://dx.doi.org/10.1029/94JA02067}{\JournalTitle{Journal of
  Geophysical Research}, 99, 23}

\bibitem[{{Gary} {et~al.}(1999{\natexlab{a}}){Gary}, {Skoug}, \&
  {Daughton}}]{Gary1999}
{Gary}, S.~P., {Skoug}, R.~M., \& {Daughton}, W. 1999{\natexlab{a}},
  \href{http://dx.doi.org/10.1063/1.873532}{\JournalTitle{Physics of Plasmas},
  6, 2607}

\bibitem[{{Gary} {et~al.}(1999{\natexlab{b}}){Gary}, {Skoug}, \&
  {Daughton}}]{Gary1999heatflux}
---. 1999{\natexlab{b}},
  \href{http://dx.doi.org/10.1063/1.873532}{\JournalTitle{Physics of Plasmas},
  6, 2607}

\bibitem[{{Graham} {et~al.}(2017){Graham}, {Rae}, {Owen}, {Walsh}, {Arridge},
  {Gilbert}, {Lewis}, {Jones}, {Forsyth}, {Coates}, \& {Waite}}]{Graham2017}
{Graham}, G.~A., {Rae}, I.~J., {Owen}, C.~J., {et~al.} 2017,
  \href{http://dx.doi.org/10.1002/2016JA023656}{\JournalTitle{Journal of
  Geophysical Research (Space Physics)}, 122, 3858}

\bibitem[{{Hammond} {et~al.}(1996){Hammond}, {Feldman}, {McComas}, {Phillips},
  \& {Forsyth}}]{Hammond1996}
{Hammond}, C.~M., {Feldman}, W.~C., {McComas}, D.~J., {Phillips}, J.~L., \&
  {Forsyth}, R.~J. 1996, \JournalTitle{Astronomy and Astrophysics}, 316, 350

\bibitem[{Hunter(2007)}]{Hunter2007}
Hunter, J.~D. 2007,
  \href{http://dx.doi.org/10.1109/MCSE.2007.55}{\JournalTitle{Computing in
  Science \& Engineering}, 9, 90}

\bibitem[{{Kajdi{\v c}} {et~al.}(2016){Kajdi{\v c}}, {Alexandrova},
  {Maksimovic}, {Lacombe}, \& {Fazakerley}}]{Kajdic2016}
{Kajdi{\v c}}, P., {Alexandrova}, O., {Maksimovic}, M., {Lacombe}, C., \&
  {Fazakerley}, A.~N. 2016,
  \href{http://dx.doi.org/10.3847/1538-4357/833/2/172}{\JournalTitle{The
  Astrophysical Journal}, 833, 172}

\bibitem[{{Kruparova} {et~al.}(2013){Kruparova}, {Maksimovic}, {{\v
  S}afr{\'a}nkov{\'a}}, {N{\v e}Me{\v c}Ek}, {Santolik}, \&
  {Krupar}}]{Kruparova2013}
{Kruparova}, O., {Maksimovic}, M., {{\v S}afr{\'a}nkov{\'a}}, J., {et~al.}
  2013, \href{http://dx.doi.org/10.1002/jgra.50468}{\JournalTitle{Journal of
  Geophysical Research (Space Physics)}, 118, 4793}

\bibitem[{{Lacombe} {et~al.}(2017){Lacombe}, {Alexandrova}, \&
  {Matteini}}]{Lacombe2017}
{Lacombe}, C., {Alexandrova}, O., \& {Matteini}, L. 2017,
  \href{http://dx.doi.org/10.3847/1538-4357/aa8c06}{\JournalTitle{The
  Astrophysical Journal}, 848, 45}

\bibitem[{{Lacombe} {et~al.}(2014){Lacombe}, {Alexandrova}, {Matteini},
  {Santol{\'{\i}}k}, {Cornilleau-Wehrlin}, {Mangeney}, {de Conchy}, \&
  {Maksimovic}}]{Lacombe2014}
{Lacombe}, C., {Alexandrova}, O., {Matteini}, L., {et~al.} 2014,
  \href{http://dx.doi.org/10.1088/0004-637X/796/1/5}{\JournalTitle{The
  Astrophysical Journal}, 796, 5}

\bibitem[{{Lazar} {et~al.}(2018){Lazar}, {Shaaban}, {Fichtner}, \&
  {Poedts}}]{Lazar2018}
{Lazar}, M., {Shaaban}, S.~M., {Fichtner}, H., \& {Poedts}, S. 2018,
  \href{http://dx.doi.org/10.1063/1.5016261}{\JournalTitle{Physics of Plasmas},
  25, 022902}

\bibitem[{{Maksimovic} {et~al.}(2005){Maksimovic}, {Zouganelis}, {Chaufray},
  {Issautier}, {Scime}, {Littleton}, {Marsch}, {McComas}, {Salem}, {Lin}, \&
  {Elliott}}]{Maksimovic2005}
{Maksimovic}, M., {Zouganelis}, I., {Chaufray}, J.-Y., {et~al.} 2005,
  \href{http://dx.doi.org/10.1029/2005JA011119}{\JournalTitle{Journal of
  Geophysical Research (Space Physics)}, 110, A09104}

\bibitem[{{Moullard} {et~al.}(2001){Moullard}, {Burgess}, {Salem}, {Mangeney},
  {Larson}, \& {Bale}}]{Moullard2001}
{Moullard}, O., {Burgess}, D., {Salem}, C., {et~al.} 2001,
  \href{http://dx.doi.org/10.1029/2000JA900144}{\JournalTitle{Journal of
  Geophysical Research (Space Physics)}, 106, 8301}

\bibitem[{{Neubauer} {et~al.}(1977){Neubauer}, {Musmann}, \&
  {Dehmel}}]{Neubauer1977}
{Neubauer}, F.~M., {Musmann}, G., \& {Dehmel}, G. 1977,
  \href{http://dx.doi.org/10.1029/JA082i022p03201}{\JournalTitle{Journal of
  Geophysical Research (Space Physics)}, 82, 3201}

\bibitem[{{Owens} {et~al.}(2017){Owens}, {Lockwood}, {Riley}, \&
  {Linker}}]{Owens2017}
{Owens}, M.~J., {Lockwood}, M., {Riley}, P., \& {Linker}, J. 2017,
  \href{http://dx.doi.org/10.1002/2017JA024631}{\JournalTitle{Journal of
  Geophysical Research (Space Physics)}, 122, 10,980}

\bibitem[{{Roberg-Clark} {et~al.}(2018){Roberg-Clark}, {Drake}, {Swisdak}, \&
  {Reynolds}}]{Roberg2018}
{Roberg-Clark}, G.~T., {Drake}, J.~F., {Swisdak}, M., \& {Reynolds}, C.~S.
  2018, \href{http://dx.doi.org/10.3847/1538-4357/aae393}{\JournalTitle{The
  Astrophysical Journal}, 867, 154}

\bibitem[{{Roberts} {et~al.}(2017){Roberts}, {Alexandrova}, {Kajdi{\v c}},
  {Turc}, {Perrone}, {Escoubet}, \& {Walsh}}]{Roberts2017}
{Roberts}, O.~W., {Alexandrova}, O., {Kajdi{\v c}}, P., {et~al.} 2017,
  \href{http://dx.doi.org/10.3847/1538-4357/aa93e5}{\JournalTitle{The
  Astrophysical Journal}, 850, 120}

\bibitem[{{Stansby} {et~al.}(2016){Stansby}, {Horbury}, {Chen}, \&
  {Matteini}}]{Stansby2016}
{Stansby}, D., {Horbury}, T.~S., {Chen}, C.~H.~K., \& {Matteini}, L. 2016,
  \href{http://dx.doi.org/10.3847/2041-8205/829/1/L16}{\JournalTitle{The
  Astrophysical Journal Letters}, 829, L16}

\bibitem[{{Stenzel}(1999)}]{Stenzel1999}
{Stenzel}, R.~L. 1999,
  \href{http://dx.doi.org/10.1029/1998JA900120}{\JournalTitle{Journal of
  Geophysical Research (Space Physics)}, 104, 14379}

\bibitem[{{Tang} {et~al.}(2020){Tang}, {Zank}, \& {Kolobov}}]{Tang2020}
{Tang}, B., {Zank}, G.~P., \& {Kolobov}, V.~I. 2020,
  \href{http://dx.doi.org/10.3847/1538-4357/ab7a93}{\JournalTitle{\apj}, 892,
  95}

\bibitem[{{Tong} {et~al.}(2019{\natexlab{a}}){Tong}, {Vasko}, {Artemyev},
  {Bale}, \& {Mozer}}]{Tong2019stasticalstudy}
{Tong}, Y., {Vasko}, I.~Y., {Artemyev}, A.~V., {Bale}, S.~D., \& {Mozer}, F.~S.
  2019{\natexlab{a}},
  \href{http://dx.doi.org/10.3847/1538-4357/ab1f05}{\JournalTitle{The
  Astrophysical Journal}, 878, 41}

\bibitem[{{Tong} {et~al.}(2019{\natexlab{b}}){Tong}, {Vasko}, {Pulupa},
  {Mozer}, {Bale}, {Artemyev}, \& {Krasnoselskikh}}]{Tong2019}
{Tong}, Y., {Vasko}, I.~Y., {Pulupa}, M., {et~al.} 2019{\natexlab{b}},
  \href{http://dx.doi.org/10.3847/2041-8213/aaf734}{\JournalTitle{The
  Astrophysical Journal Letters}, 870, L6}

\bibitem[{{{\v S}tver{\'a}k} {et~al.}(2009){{\v S}tver{\'a}k}, {Maksimovic},
  {Tr{\'a}vn{\'{\i}}{\v c}ek}, {Marsch}, {Fazakerley}, \&
  {Scime}}]{Stverak2009}
{{\v S}tver{\'a}k}, {\v S}., {Maksimovic}, M., {Tr{\'a}vn{\'{\i}}{\v c}ek},
  P.~M., {et~al.} 2009,
  \href{http://dx.doi.org/10.1029/2008JA013883}{\JournalTitle{Journal of
  Geophysical Research (Space Physics)}, 114, A05104}

\bibitem[{{{\v S}tver{\'a}k} {et~al.}(2008){{\v S}tver{\'a}k},
  {Tr{\'a}vn{\'{\i}}{\v c}ek}, {Maksimovic}, {Marsch}, {Fazakerley}, \&
  {Scime}}]{stverak2008}
{{\v S}tver{\'a}k}, {\v S}., {Tr{\'a}vn{\'{\i}}{\v c}ek}, P., {Maksimovic}, M.,
  {et~al.} 2008,
  \href{http://dx.doi.org/10.1029/2007JA012733}{\JournalTitle{Journal of
  Geophysical Research (Space Physics)}, 113, A03103}

\bibitem[{{{\v S}tver{\'a}k} {et~al.}(2015){{\v S}tver{\'a}k},
  {Tr{\'a}vn{\'{\i}}{\v c}ek}, \& {Hellinger}}]{Stverak2015}
{{\v S}tver{\'a}k}, {\v S}., {Tr{\'a}vn{\'{\i}}{\v c}ek}, P.~M., \&
  {Hellinger}, P. 2015,
  \href{http://dx.doi.org/10.1002/2015JA021368}{\JournalTitle{Journal of
  Geophysical Research (Space Physics)}, 120, 8177}

\bibitem[{{Vasko} {et~al.}(2019){Vasko}, {Krasnoselskikh}, {Tong}, {Bale},
  {Bonnell}, \& {Mozer}}]{Vasko2019}
{Vasko}, I.~Y., {Krasnoselskikh}, V., {Tong}, Y., {et~al.} 2019,
  \href{http://dx.doi.org/10.3847/2041-8213/ab01bd}{\JournalTitle{The
  Astrophysical Journal Letters}, 871, L29}

\bibitem[{{Veltri} \& {Zimbardo}(1993)}]{Veltri1993}
{Veltri}, P., \& {Zimbardo}, G. 1993,
  \href{http://dx.doi.org/10.1029/93JA00812}{\JournalTitle{Journal of
  Geophysical Research (Space Physics)}, 98, 13325}

\bibitem[{{Vocks}(2012)}]{Vocks2012}
{Vocks}, C. 2012,
  \href{http://dx.doi.org/10.1007/s11214-011-9749-0}{\JournalTitle{Space
  Science Reviews}, 172, 303}

\bibitem[{{Vocks} \& {Mann}(2003)}]{Vocks2003}
{Vocks}, C., \& {Mann}, G. 2003,
  \href{http://dx.doi.org/10.1086/376682}{\JournalTitle{The Astrophysical
  Journal}, 593, 1134}

\bibitem[{{Vocks} {et~al.}(2005){Vocks}, {Salem}, {Lin}, \& {Mann}}]{Vocks2005}
{Vocks}, C., {Salem}, C., {Lin}, R.~P., \& {Mann}, G. 2005,
  \href{http://dx.doi.org/10.1086/430119}{\JournalTitle{The Astrophysical
  Journal}, 627, 540}

\bibitem[{{Wilson} {et~al.}(2013){Wilson}, {Koval}, {Szabo}, {Breneman},
  {Cattell}, {Goetz}, {Kellogg}, {Kersten}, {Kasper}, {Maruca}, \&
  {Pulupa}}]{Wilson2013}
{Wilson}, L.~B., {Koval}, A., {Szabo}, A., {et~al.} 2013,
  \href{http://dx.doi.org/10.1029/2012JA018167}{\JournalTitle{Journal of
  Geophysical Research (Space Physics)}, 118, 5}

\bibitem[{{Zhang} {et~al.}(1998){Zhang}, {Matsumoto}, \& {Kojima}}]{Zhang1998}
{Zhang}, Y., {Matsumoto}, H., \& {Kojima}, H. 1998,
  \href{http://dx.doi.org/10.1029/98JA01371}{\JournalTitle{Journal of
  Geophysical Research (Space Physics)}, 103, 20529}

\end{thebibliography}

\clearpage

\end{document}